\documentclass[prc,twocolumn,showpacs,twoside,floatfix]{revtex4}
\usepackage{graphicx}
\usepackage{amssymb}
\usepackage{amsmath}

\begin{document}

\title{Three-body decay of $^{6}$Be}
\author{L.~V.~Grigorenko$^{1,2,3}$, T.~D.~Wiser$^{4}$,
K.~Mercurio$^{4}$, R.~J.~Charity$^{5}$, R.~Shane$^{4},$ L.~G.~Sobotka$^{4,5}$, 
J.~M.~Elson$^{5}$, A.~Wuosmaa$^{6}$, A.~Banu$^{7}$, M. McCleskey$^{7}$,
L.~Trache$^{7}$, R.~E.~Tribble$^{7}$, and M.~V.~Zhukov$^{8}$}

\affiliation{$^{1}$Flerov Laboratory of Nuclear Reactions, JINR, RU-141980 
Dubna, Russia. \\
$^{2}$Gesellschaft f\"{u}r Schwerionenforschung mbH, Plankstrasse 1,
D-64291, Darmstadt, Germany.\\
$^{3}$RRC ``The Kurchatov Institute'', Kurchatov sq.~1, 123182 Moscow,
Russia.\\
Departments of Physics$^{4}$ and Chemistry$^{5}$, Washington University,
St.~Louis, Missouri 63130, USA. \\
$^6$Department of Physics, Western Michigan University, Kalamazoo, Michigan
49008, USA. \\
$^7$Cyclotron Institute, Texas A\&M University, College Station, Texas
77843, USA. \\
$^{8}$Fundamental Physics, Chalmers University of Technology, S-41296 
G\"{o}teborg, Sweden.}


\begin{abstract}
Three-body correlations for the ground-state decay of the lightest
two-proton emitter $^{6}$Be are studied both theoretically and
experimentally. Theoretical studies are performed in a three-body
hyperspherical-harmonics cluster model. In the experimental studies, the
ground state of $^{6}$Be was formed following the $\alpha $ decay of a $^{10}$C 
beam inelastically excited through interactions with Be and C targets.
Excellent agreement between theory and experiment is obtained demonstrating
the existence of complicated correlation patterns which can elucidate the
structure of $^{6}$Be and, possibly, of the $A$=6 isobar.
\end{abstract}

\maketitle


\section{Introduction}

\begin{figure}[tbp]
\includegraphics[width=0.33\textwidth]{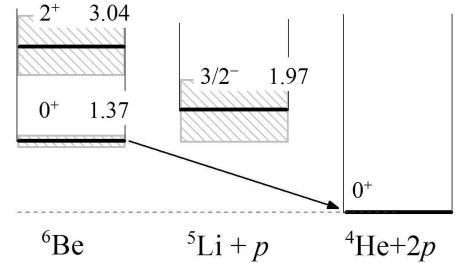}
\caption{Energy levels and decay scheme for $^{6}$Be \protect\cite{til02}.
The $^{6}$Be g.s.\ is a true two-proton emitter in the sense of Goldansky:
the sequential decay of this state is not possible as the lowest possible
intermediate, $^{5}$Li g.s., is not energetically accessible.}
\label{fig:6be-spec}
\end{figure}

Two-proton (2\textit{p}) radioactivity was predicted by V.I. Goldansky in 1960 
\cite{gol60} as an exclusively quantum-mechanical phenomenon. True three-body 
decay, in his terms, is a situation where the sequential emission of the 
particles is energetically prohibited and all the final-state fragments are 
emitted simultaneously. These energy conditions are illustrated for $^6$Be in 
Fig.~\ref{fig:6be-spec} which shows that the $^{5}$Li ground state (g.s.) is not 
fully accessible for sequential decay. Since the experimental discovery of 
$^{45}$Fe two-proton radioactivity in 2002 \cite{pfu02,gio02}, this field has 
made fast progress. New cases of 2\textit{p} radioactivity were found for 
$^{54}$Zn \cite{bla05}, $^{19}$Mg \cite{muk07}, and, maybe, $^{48}$Ni 
\cite{dos05}. The 2\textit{p} correlations were recently measured in $^{45}$Fe 
\cite{mie07}, $^{19}$Mg \cite{muk07,muk08}, $^{16}$Ne \cite{muk08}, and $^{10}$C 
\cite{Mercurio08}. Very interesting \cite{muk06}, but, so far, controversial 
\cite{pec07} case is possibility of $2p$ emission off deformed isomeric state in 
$^{94}$Ag. All these decays exhibit complex correlation patterns. It is argued 
that studies of these patterns could provide important information about the 
structure of the decaying nuclei.

With this active research as the background, there is one case which has been 
unduly forgotten. The $^{6}$Be nucleus is the lightest true two-proton emitter 
in the sense of Goldansky. As this is expected to be the simplest case (smallest 
Coulomb interaction, expressed cluster structure with closed-shell core), a full 
understanding of its physics would provide a reliable basis for all future 
studies of 2\textit{p} decay. However until now, theoretical work on $^{6}$Be 
was limited to predicting the energies and widths of its states. In addition, 
precise experimental data do not exist. The last experimental work dedicated to 
correlations in $^{6}$Be g.s.\ is about 20 years old.

There is one more aspect which makes the $^{6}$Be case of special interest.
In the last decade, large efforts have been directed to studies of $^{6}$He
with special emphasis to the understanding of the halo properties in this
comparatively simple and accessible case. The associated literature
comprises hundreds of titles. To deduce the correlations in the neutron
halo,\ one has to excite (e.g., Coulex) or destroy (e.g., knockout
reactions) this nucleus. Therefore, the interpretation of the experimental
data is influenced by the need to clarify details of the reaction mechanism
\footnote{Evident exception is, of cause, the $\beta $-decay of $^{6}$He and 
$\beta $-delayed particle emission. These reactions exploit very ``reliable'' 
weak probe, providing important, but naturally limited information about this 
system.}. However, information about mirror system can be obtained without all 
this ``violence''. The isobaric analogue state in $^{6}$Be decays to the 
$\alpha$+$p$+$p$ channel all by itself, providing the differential data on 
correlations. This data can be used directly to elucidate the structure of 
$^{6}$Be without the need to deal with the details of the reaction mechanism. 
Thus an important opportunity exists for a better understanding of $^{6}$He 
properties through detailed studies of the $^{6}$Be. This has not been exploited 
previously.

In this work, we provide detailed theoretical calculations of the three-body 
decay characteristics of $^{6}$Be in a three-body cluster $\alpha $+$p$+$p$ 
model. We demonstrate that, in certain aspects, $^{6}$Be may be a preferable 
tool for studies of the $A$=6 isobar, especially considering the high 
sensitivity of observables to the details of the theoretical models. We then 
discuss previous experimental and theoretical works on $^6$Be. Subsequently, we 
report on an experiment where $^{6}$Be fragments are formed after the $\alpha 
$-decay of $^{10}$C projectiles excited by inelastic scattering. These data 
cover the complete kinematic space accessible for three-body decay and the 
correlations are compared to the theoretical predictions.

The $\hbar=c=1$ system of units is used in this work. The following notations 
are used: $E_T$ is the system energy and $E_{3r}$ is the three-body resonance 
energy relative to the three-body $\alpha$+$p$+$p$ threshold.


\section{Theoretical model}


The theoretical framework of this paper is largely the same as that developed 
for the two-proton radioactivity and three-body decay studies in 
Refs.~\cite{gri02,gri02a,gri03,gri03a,gri07,gri07a}. It is based on the 
three-body cluster model using the hyperspherical-harmonics (HH) method. The 
predictions obtained with this approach were found to be in very good agreement 
with experimental widths and momentum distributions \cite{mie07,muk07,muk08}.

In this section, we sketch the necessary formalism emphasizing only the
points which differ from previous treatments.


\subsection{Hyperspherical harmonics method}

\begin{figure}[tbp]
\includegraphics[width=0.48\textwidth]{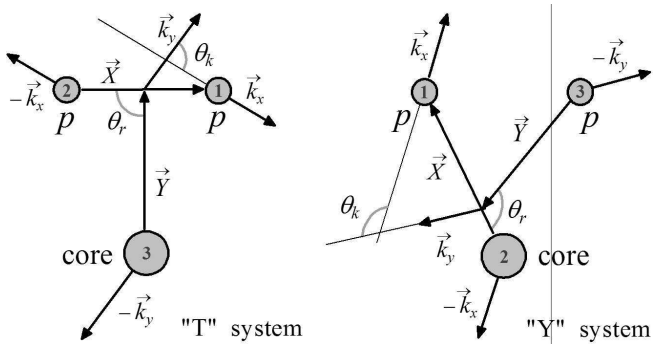}
\caption{Independent ``T'' and ``Y'' Jacobi systems for the core+$N$+$N$
three-body system in coordinate and momentum spaces.}
\label{fig:jacobi}
\end{figure}

For narrow states, the time-dependent wavefunction (WF) in a finite domain
can be parameterized as
\begin{equation}
\Psi _{3}^{(+)}(\rho ,\Omega _{\rho },t)=e^{-\frac{\Gamma }{2}t-iEt}\;\Psi
_{3}^{(+)}(\rho ,\Omega _{\rho })\;.
\label{eq:decay-param}
\end{equation}
The radial part of this WF can be obtained with good precision as a solution
of the inhomogeneous system of equations
\begin{subequations}
\label{eq:source}
\begin{gather}
\left( \hat{H}-E_{3r}\right) \Psi _{E}^{(+)}(\rho ,\Omega _{\rho })
=-i\,(\Gamma /2)\,\Psi _{\text{box}}(\rho ,\Omega _{\rho })\;, \\
\hat{H} 
=\hat{T}+\hat{V}_{cp}(\mathbf{r}_{cn_{1}})+\hat{V}_{cp}(\mathbf{r}_{cp_{2}}) 
+\hat{V}_{pp} (\mathbf{r}_{p_{1}p_{2}})\;.
\end{gather}
\end{subequations}
Here $\Psi _{\text{box}}$ and $E_{3r}$ are the eigenfunction and the
eigenvalue of the equation
\begin{equation}
\left( \hat{H}-E\right) \Psi _{\text{box}}(\rho ,\Omega _{\rho })=0\;,
\label{eq:box}
\end{equation}
solved with a ``box'' boundary condition at large $\rho $. The hyperspherical 
coordinates are defined via the Jacobi
vectors
\begin{subequations}
\label{eq:Jacobi}
\begin{eqnarray}
\mathbf{X} &=&\mathbf{r}_{p_1}-\mathbf{r}_{p_2}\;,\quad 
\mathbf{Y}=(\mathbf{r}_{p_1}+\mathbf{r}_{p_2})/2-\mathbf{r}_{c}\;, \\
\rho ^{2} &=&\frac{2}{3}\left( r_{cp_{1}}^{2}+r_{cp_{2}}^{2}\right) 
+\frac{1}{6}\,r_{p_{1}p_{2}}^{2}=\frac{1}{2}\,X^{2}+\frac{4}{3}\,Y^{2}\;, \\
\Omega _{\rho } &=&\{\theta _{\rho },\Omega _{x},\Omega _{y}\}\;,\quad
\theta _{\rho }=\text{arctan}\left[ \sqrt{\frac{3}{8}}\frac{X}{Y}\right] \;.
\end{eqnarray}
\end{subequations}
These Jacobi variables are given in ``T'' Jacobi system (see 
Fig.~\ref{fig:jacobi}). The hyperradial components $\chi _{K\gamma }^{(+)}(\rho 
)$ of the WF equation\ \ref{eq:source}, possessing the pure outgoing asymptotics
\begin{equation}
\Psi _{E}^{(+)}(\rho ,\Omega _{\rho })=\rho ^{-5/2}\sum_{K\gamma }^{K_{\max
}}\,\chi _{K\gamma }^{(+)}(\varkappa \rho )\,\mathcal{J}_{K\gamma
}^{JM}(\Omega _{\rho })\;,
\end{equation}
are matched to approximate boundary conditions of the three-body Coulomb
problem obtained in Ref.~\cite{gri03c}. The radial components of this WF at
large $\rho $ values can be represented as
\begin{equation}
\chi _{K\gamma }^{(+)}(\varkappa \rho )\sim \, A_{KLl_{x}l_{y}} ^{JSS_{x}} 
(\varkappa ) \;\tilde{\mathcal{H}}_{K \gamma}^{(+)}(\varkappa \rho )\,.
\label{eq:psi3plus-ass}
\end{equation}
In general, the functions $\tilde{\mathcal{H}}_{K\gamma }^{(+)}$ are some
linear combinations of Coulomb functions with the outgoing asymptotic $G+iF$. 
The functions $\mathcal{J}_{K\gamma
}^{JM}(\Omega _{\rho })$ are hyperspherical harmonics coupled with spin 
functions to total spin $J$. ``Multyindex'' $\gamma$ denote the complete set of 
quantum numbers except the principal quantum number $K$: 
$\gamma=\{L,S,l_x,l_y\}$. The value $K_{\max }$ truncates the hyperspherical 
expansion. The
hypermoment $\varkappa $ is expressed via the energies of the subsystems 
$E_{x}$, $E_{y}$ or via the Jacobi momenta $k_{x}$, $k_{y}$ conjugate to
Jacobi coordinates $X$, $Y$:
\begin{subequations}
\label{eq:Jmom}
\begin{eqnarray}
\mathbf{k}_{x} & = & \frac{1}{2}\left( \mathbf{k}_{p_{1}} 
-\mathbf{k}_{p_{2}}\right), \\
\mathbf{k}_{y} & = & \frac{2}{3}\left( \mathbf{k}_{p_{1}} 
+\mathbf{k}_{p_{2}}\right) -\frac{1}{3}\mathbf{k}_{c}, \\
\varkappa ^{2} & = & 2ME_T=2M(E_{x}+E_{y})=2k_{x}^{2}+\frac{3}{4}\,k_{y}^{2}\;, 
\\
\Omega _{\varkappa } & = & \{\theta _{k},\Omega _{k_{x}},\Omega
_{k_{y}}\}\;,\quad \theta _{k}=\text{arctan}[E_{x}/E_{y}]\;.
\end{eqnarray}
\end{subequations}
A more detailed picture of the ``T'' and ``Y'' Jacobi systems in coordinate and 
momentum spaces can be found in Fig.~\ref{fig:jacobi}.

The set of coupled equations for the functions $\chi ^{(+)}$ has the form
\begin{multline}
\left[ \frac{d^{2}}{d\rho ^{2}} - \frac{\mathcal{L}(\mathcal{L}+1)}{\rho ^{2}}
+2M\left\{ E-V_{K\gamma ,K\gamma }(\rho )\right\} \right] \chi _{K\gamma
}^{(+)}(\rho )= \\
2M\sum_{K^{\prime }\gamma ^{\prime }}V_{K\gamma ,K^{\prime }\gamma ^{\prime
}}(\rho )\chi _{K^{\prime }\gamma ^{\prime }}^{(+)}(\rho )+i\,\Gamma M\,\chi
_{K\gamma }(\rho )\,,  \label{shredl}
\end{multline}
where $\mathcal{L}=K+3/2$ is ``effective angular momentum'' and $V_{K\gamma 
,K^{\prime }\gamma ^{\prime
}}(\rho )$ is ``three-body potential'' (matrix elements of the pairwise
potentials);
\begin{multline}
V_{K\gamma ,K^{\prime }\gamma ^{\prime }}(\rho ) = \\
\int \!\!d\Omega _{\rho }\,\mathcal{J}_{K^{\prime }\gamma ^{\prime
}}^{JM\ast }(\Omega _{\rho 
})\sum_{i<j}V_{ij}(\mathbf{r}_{ij})\,\mathcal{J}_{K\gamma }^{JM}(\Omega _{\rho 
})\,  \label{hhpot}
\end{multline}
and
\begin{equation}
\Psi _{\text{box}}(\rho ,\Omega _{\rho }) =\rho ^{-5/2}\sum_{K\gamma }\chi
_{K\gamma }(\rho )\,\mathcal{J}_{K\gamma }^{JM}(\Omega _{\rho })\,.
\end{equation}

\subsection{Width and momentum distribution}


Equation~\ref{eq:source} is first solved with an arbitrary value of $\Gamma $
and then the width is found according to the ``natural'' definition as the
flux $j$ through a hypersphere with large radius $\rho _{\max }$ divided by the
internal normalization $N$ (``number of particles'' inside the sphere):
\begin{eqnarray}
\Gamma _{\text{nat}} & = & j/N\;,  \label{eq:nat} \\
j & = & \int d\Omega _{\rho }\,\frac{d\,j(\rho _{\max },\Omega _{\rho })}{d 
\,\Omega _{\rho }}\;,
\label{eq:tot-flux} \\
N & = & \sum_{K\gamma }N_{K\gamma }=\sum_{K\gamma }\int_{0}^{\rho _{\text{int}}} 
\!d\rho \,\left\vert \chi _{K\gamma }^{(+)}(\rho )\right\vert ^{2}\;.
\label{eq:int-norm}
\end{eqnarray}
The differential flux through the hypersphere $\rho _{\max }$ is defined as
\begin{multline}
\frac{dj(\rho _{\max },\Omega _{\rho })}{d\Omega _{\rho }}= \\
\left.\mathop{\rm Im}\Bigl[\;\Psi _{3}^{(+)\dagger }\,\rho ^{5/2}\,\frac{d}{M 
d\rho }\,\rho ^{5/2}\,\Psi _{3}^{(+)}\Bigr]\right\vert _{\rho =\rho _{\max
}}\,.  \label{eq:dif-flux}
\end{multline}
If, for sufficiently large $\rho $, the coefficients $A_{Ll_{x}l_{y}}^{KSS_{x}}$ 
in Eq.~\ref{eq:psi3plus-ass} become independent
of $\rho $, then the coordinate distribution becomes identical to the
momentum distribution, i.e.,
\begin{equation}
\frac{j(\rho _{\max },\Omega _{\rho })}{d\Omega _{\rho }}\,\rightarrow \frac{
dj(\Omega _{\varkappa })} {d\Omega _{\varkappa }}\;.
\label{eq:current-equiv}
\end{equation}
Further discussions of the validity of this approximation 
(Eq.~\ref{eq:current-equiv}), and detailed expressions for the momentum
distributions, can be found in Ref.~\cite{gri03c}.


\subsection{Potentials}


The $NN$ potential is taken either as a simple $s$-wave single-Gaussian form
BJ (from the book of Brown and Jackson \cite{bro76})
\begin{equation}
V(r)=V_{0}\exp (-r^{2}/r_{0}^{2})\;,
\end{equation}
with $V_{0}=-31$ MeV and $r_{0}=1.8$ fm, or the realistic ``soft-core''
potential GPT (Gogny-Pires-de Tourreil \cite{gog70}).

The Coulomb potential of the homogeneously charged sphere $r_{\text{sph}}=1.852$ 
fm is used in the $\alpha $-$p$ channel. In addition for this
channel, we use an $\ell$-dependent potential SBB (Sack-Biedenharn-Breit
\cite{sac54})
\begin{equation}
V(r)=V_{c}^{(\ell)}\exp (-r^{2}/r_{0}^{2})+(\mathbf{\ell}\cdot\mathbf{s})\,
V_{\ell s}\exp(-r^{2}/r_{0}^{2})\;,
\end{equation}
where $r_{0}=2.30$~fm, $V_{c}^{(0)}=50$~MeV, $V_{c}^{(1)}=-47.32$~MeV, 
$V_{c}^{(2)}=-23$~MeV, and $V_{ls}=-11.71$~MeV. Historically, a somewhat
modified SBBM potential has been used in the calculations of $A$=6 isobars
in order to better reproduce the binding energies (e.g., Ref.~\cite{dan91}).
Later it was realized that it is more consistent to provide the
phenomenological binding-energy correction using an additional short-range
three-body potential (see, e.g., the discussion in Ref.~\cite{gri07}). In this
work, we used a short-range three-body potential of the form
\begin{equation}
V_{3}(\rho )=\delta _{K\gamma ,K^{\prime }\gamma ^{\prime
}}V_{3}^{(0)}/[1+\exp ((\rho -\rho _{0})/d_{3})]\;,  \label{eq:pot3}
\end{equation}
where $\rho _{0}=2.5$~fm and $d_{3}=0.4$~fm. This ``short-range'' three-body 
potential (note the small diffuseness) does not distort the interactions in the
subbarrier region which was found to be important for consistent studies of 
decay properties.

Three sets of nuclear potential are employed in this work. They are denoted as 
P1 (SBB+BJ), P2 (SBB+GPT), and P3 (SBBM+GPT). The values of $V_{3}^{(0)}$ used 
with potential sets P1, P2, and P3 are $-11.14$~MeV, $-13.22$~MeV, and 
$-0.64$~MeV, respectively. Throughout this paper when the potential set is not 
specified, the results of the calculations with the P2 set are shown.


\subsection{Reaction models}

\label{sec:reac}

\begin{figure}[tbp]
\includegraphics[width=0.45\textwidth]{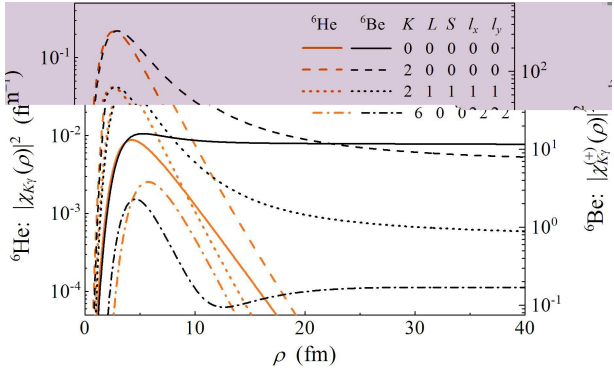}
\caption{Densities $|\protect\chi_{K\protect\gamma}(\protect\rho)|^2$ and $|%
\protect\chi^{(+)}_{K\protect\gamma}(\protect\rho)|^2$ for the largest
components of the $^{6}$He and $^{6}$Be g.s.\ WFs.}
\label{fig:wfs}
\end{figure}

In general, different definitions for the width of a decaying state coincide
only in the limit when the width is very small. For the ground state of $^{6}
$Be, the definition of Eq.~\ref{eq:nat} is not very precise, as this state
is comparatively broad ($\Gamma =92\pm 6$ keV) and thus the internal
normalization $N$ (Eq.~\ref{eq:int-norm}) is sensitive to the integration
limit $\rho _{\text{int}}$. For reasonable values of $\rho _{\text{int}}$
ranging from $10-20$~fm, the uncertainty in the width ($\Gamma _{\text{nat}}$)
is about $25\%$ [see Sec.~\ref{sec:com-modis}, Fig.~\ref{fig:corel-rho-dep}(a)]. 
This problem does not exist for narrow $2p$ emitters ($\Gamma <1$ eV) where the 
WFs $\chi _{K\gamma }^{(+)}$ are vanishingly small under the Coulomb barrier. 
The densities for the dominating components of the $^{6}$He and $^{6}$Be WFs are 
shown in Fig.\ \ref{fig:wfs}. For $^{6}$Be, it is clear that the WF under the 
barrier is not negligible.

For moderately broad states, there are alternative ways to derive the width. 
These involve either the study of the $3\rightarrow 3$ scattering or the study 
of a particular reaction. For technical reasons, the latter is preferable for 
our application. For example, in order to determine the population of $^{6}$Be 
in a charge-exchange reaction on $^{6}$Li at zero angle, Eq.~\ref{eq:source} can 
be reformulated as
\begin{multline}
\left( \hat{H}-E_{T}\right) \Psi _{^{6}\text{Be}}^{(+)}(\rho ,\Omega _{\rho
})= \\
\sum_{i}\tau _{i}^{-}\sum_{M}\sigma _{i}^{(M)}\;\Psi _{^{6}\text{Li}}^{JM} (\rho 
,\Omega _{\rho })\;.
\label{eq:ste-6li}
\end{multline}
This notation is based on the fact that for angles close to zero, the
transitions in charge-exchange reactions, in the limit of high energies, are
provided by the Gamow-Teller operator. Although this reaction is different
to the one studied experimentally in this work, it is sufficient for our
computational purposes. Namely, we will demonstrate that for the $^{6}$Be
g.s.\ population, the choice of the reaction mechanism is not very important
(there are still some exclusive situations, which we will discuss elsewhere).

Using the source function of Eq.~\ref{eq:ste-6li}, the cross section for the
population of the three-body continuum is proportional to the outgoing flux
of the three particles on a hypersphere of some large radius $\rho = \rho_{\max 
}$:
\begin{equation}
d\sigma (E_{T})/dE_{T}\sim j(\rho _{\max },\Omega _{\rho })\,.
\label{eq:cross}
\end{equation}
Differentials of this flux on the hypersphere provide angular and energy
distributions among the decay products at the given decay energy $E_{T}$ in
analogy with Eqs.~\ref{eq:dif-flux} and \ref{eq:current-equiv}.


\subsection{``Feshbach'' reduction}


Although the HH calculations for $^{6}$Be can be performed with $K_{\max
}=22-26$, these basis sizes may not be sufficient to obtain good convergence
for all observables. However, the basis size can be effectively increased
using the adiabatic procedure based on the so-called Feshbach reduction (FR)
\cite{gri07}. Feshbach reduction eliminates from the total WF $\Psi =\Psi
_{p}+\Psi _{q\text{,}}$ an arbitrary subspace $q$ using the Green's function
of this subspace:
\begin{equation}
H_{p}=T_{p}+V_{p}-V_{pq}G_{q}V_{pq}\;.
\end{equation}
In an adiabatic approximation, we can assume that the radial part of
kinetic energy is small under the centrifugal barrier in the channels where
this barrier is large and can be approximated as a constant. In this
approximation, the FR procedure is reduced to the construction of effective
three-body interactions $V_{K\gamma ,K^{\prime }\gamma ^{\prime }}^{\text{eff%
}}$ by the matrix operations
\begin{multline}
G_{K\gamma ,K^{\prime }\gamma ^{\prime }}^{-1} =(H-E)_{K\gamma ,K^{\prime
}\gamma ^{\prime }}= \\
V_{K\gamma ,K^{\prime }\gamma ^{\prime }} \\
+\left[ E_{f}-E+\frac{(K+3/2)(K+5/2)}{2M\rho ^{2}}\right] \delta _{K\gamma
,K^{\prime }\gamma ^{\prime }}
\end{multline}
where
\begin{multline}
V_{K\gamma ,K^{\prime }\gamma ^{\prime }}^{\text{eff}} = \\
V_{K\gamma ,K^{\prime }\gamma ^{\prime }}-\sum V_{K\gamma ,\bar{K}\bar{\gamma%
}}G_{\bar{K}\bar{\gamma},\bar{K}^{\prime }\bar{\gamma}^{\prime }}V_{\bar{K}%
^{\prime }\bar{\gamma}^{\prime },K^{\prime }\gamma ^{\prime }}\;.
\end{multline}
Summations over indexes with the bar are made for the eliminated channels. We
typically eliminate the channels with $K>K_{FR}$, where $K_{FR}$ provides
the sector of the hyperspherical basis where the calculations remains fully
dynamical. We take the \textquotedblleft Feshbach energy\textquotedblright\ $%
E_{f}$ in our calculations as $E_{f}\equiv E$.

There are two ways to control the reliability of the FR procedure. (i) The
``soft'' method is to vary $K_{FR}$ from the maximum attainable in the
dynamic calculations downwards for fixed $K_{\max }$. The results, in
principle, should coincide. (ii) The ``safe'' method is to take $K_{\max }$
in the range attainable for dynamic calculations and compare the ``reduced''
$K_{\max }\rightarrow K_{FR}$ calculations (with much smaller dynamic basis
size $K_{FR}$) with completely dynamic calculations with $K_{\max }$. For $%
^{6}$Be, these considerations show that we can safely use $K_{FR}=14$.
However, the even safer value of $K_{FR}=22$ is used in this work.


\section{Ground state}


There are several convergence characteristics that should be understood
before reliable results on $^{6}$Be are obtained. The convergence character
is quite different for all the observables of interest and also depend
strongly on the interaction in the $p$-$p$ channel.


\subsection{Convergence of energy and width}

\begin{figure}[tbp]
\includegraphics[width=0.38\textwidth]{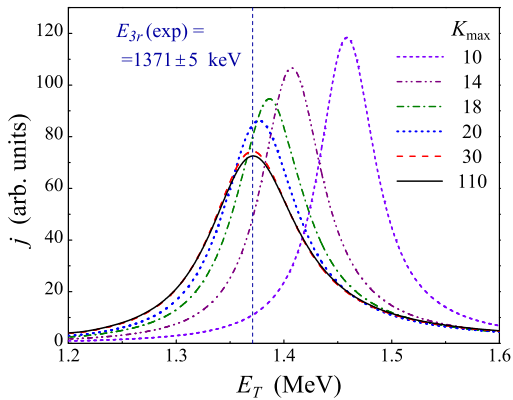}
\caption{(Color online) Energy profile of the $^{6}$Be g.s. populated in the
charge-exchange reaction with $^{6}$Li. The results are shown as a function
of the basis size $K_{\max}$ where $K_{FR}=22$. For $K_{\max}\leq K_{FR}$
and no Feshbach reduction is needed.}
\label{fig:sig-con-k}
\end{figure}

Because of the problem mentioned in Sect.~\ref{sec:reac}, we need to begin
our studies with the energy dependence of the cross section. The convergence
of the cross-section profile with increasing size of the basis is
demonstrated in Fig.~\ref{fig:sig-con-k}. The main character of the
convergence is clearly seen here; the centroid energy decreases, while the
width grows significantly.

\begin{figure}[tbp]
\includegraphics[width=0.39\textwidth]{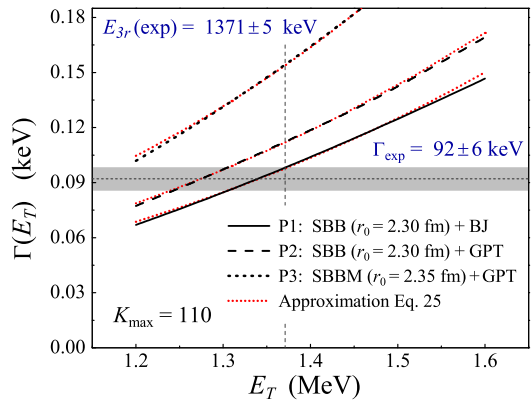}
\caption{(Color online) Dependence of $^{6}$Be g.s. width on the decay
energy $E_T$. Predications are shown for the three potential set P1-P3. The
dotted curves show the approximation of Eq.~\protect\ref{eq:g-ot-e}. }
\label{fig:pen-ot-e}
\end{figure}

The cross section for the $^{6}$Be g.s.\ population, shown in 
Fig.~\ref{fig:sig-con-k}, clearly has a profile close to a slightly asymmetric
Lorentzian. Can the profile of this three-body resonance be described by
appropriately-modified R-matrix type expressions? A curious result is
obtained here, the resonance profile, shown in Fig.~\ref{fig:sig-con-k} by
the solid curve, can be fit with amazing precision by the following
expression:
\begin{equation}
\sigma (E_{T})\sim \frac{\Gamma (E_{T})}{(E_{T}-E_{3r})^{2}+\Gamma
(E_{T})^{2}/4}\;,  \label{eq:sig-prof}
\end{equation}
\begin{equation}
\Gamma (E_{T})=\Gamma _{0}\left[ \alpha \left( \frac{E_{T}}{E_{3r}}\right)
^{2}+(1-\alpha )\left( \frac{E_{T}}{E_{3r}}\right) ^{4}\right] \;,
\label{eq:g-ot-e}
\end{equation}
where $\Gamma _{0}=98$ keV and $\alpha =0.65$. Equation~\ref{eq:sig-prof}
is the ordinary expression for the inelastic cross section of an isolated
resonance. The parameterization of Eq.~\ref{eq:g-ot-e} was chosen because, for
the single-channel penetration through the hyperspherical barrier with $K=0$, 
the energy dependence of the width can be inferred as $\Gamma (E_{T})\sim
E_{T}^{2}$. For $K=2$ one has $\Gamma (E_{T})\sim E_{T}^{4}$ (see, e.g.,
Ref.~\cite{gol04}. It should be understood that the $K=0$ component is
equivalent to a ``phase volume'' with the characteristic energy behavior of $%
\sim E^{2}$). The energy dependence of the width obtained by Eq.~\ref{eq:g-ot-e} 
almost coincides with the calculated dependence of this width
in a reasonable energy range (see Fig.~\ref{fig:pen-ot-e} when one
uses $\alpha$=0.63 and 0.52 for potential P2 and P3, respectively). If we
take the actual calculated partial widths for the $K=0$ and $2$ components
from Table \ref{tab:struc-t}, then the value of $\alpha $ can be estimated
as
\begin{equation}
\alpha =N_{K=0}/(N_{K=0}+N_{K=2})\approx 0.58 \; .
\end{equation}
This is quite close to the value 0.65 obtained by a fit.

The existence of this simple approximation, despite the fact that there are
Coulomb interactions and other numerous channels involved, may demonstrate
that the dynamics of the $^{6}$Be g.s.\ decay is largely defined by the
penetration through the hyperspherical barriers. Possibly, this is due to
the comparatively large $^{6}$Be decay energy of $E_{3r}=1.371$~MeV. Simple
estimates shows that the state is ``sitting'' somewhere straight on the top
of the Coulomb barrier.

It was found that the value of $j(E_{T})$ for $^{6}$Be g.s.\ is not
sensitive to the particular choice of the source in Eq.~\ref{eq:ste-6li},
which is typically within the width of the line \footnote{%
There are, however, some very special situations which we can not discuss in
this work.}. This means that the width defined by the procedure of
Eqs.~\ref{eq:sig-prof} and \ref{eq:g-ot-e} is very reliable. We can
fine tune the
value $\rho _{\text{int}}$ in Eq.~\ref{eq:int-norm} so that the definition
of the width in Eq.~\ref{eq:nat} coincides with the definition in Eq.~\ref%
{eq:sig-prof} and subsequently we can reliably use Eq.~\ref{eq:nat}. All of
the potential sets P1-P3 needed $\rho _{\text{int}}\approx 12.5$~fm.

\begin{figure}[tbp]
\includegraphics[width=0.44\textwidth]{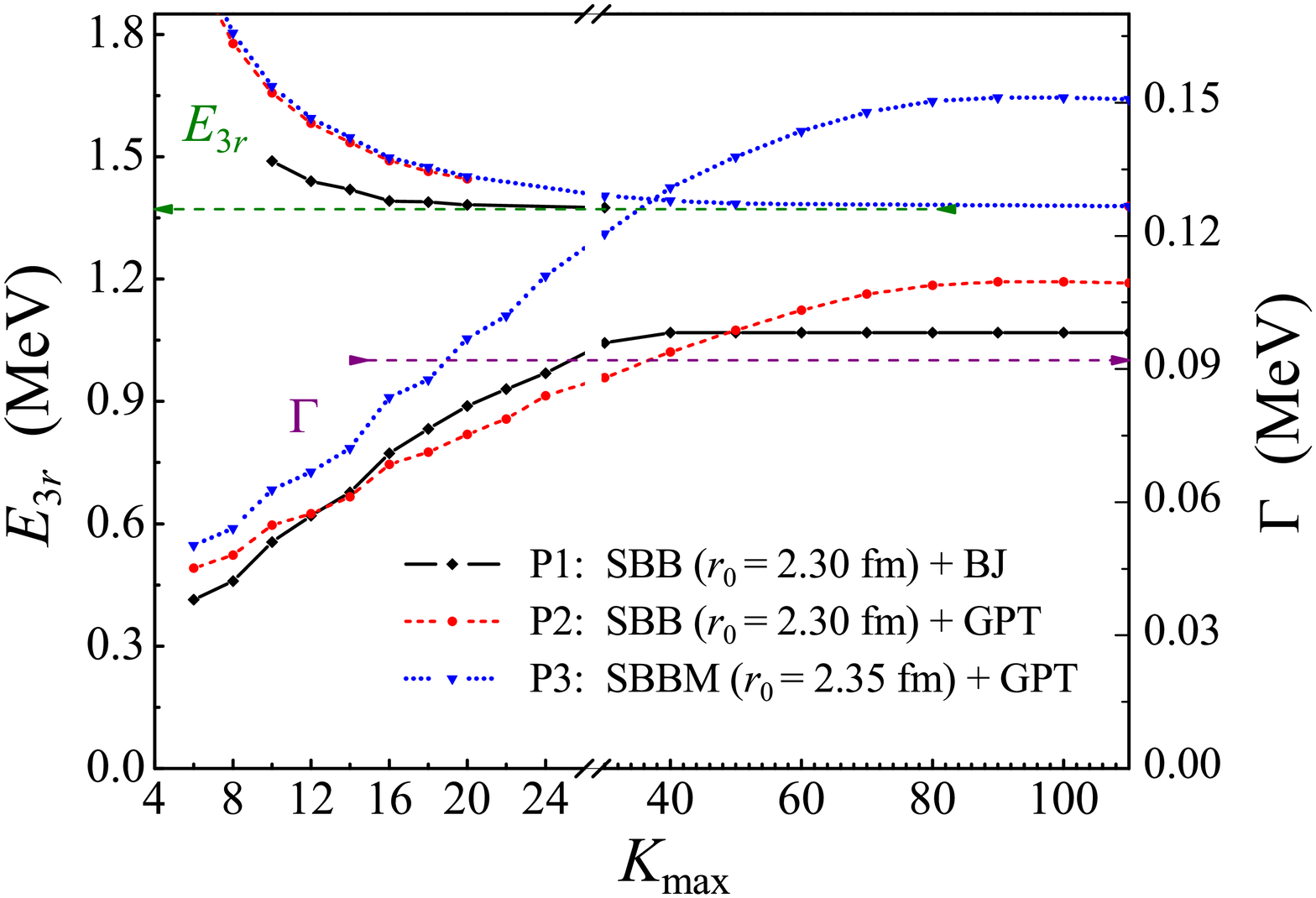}
\caption{(Color online) Convergency of the resonance energy $E_{3r}$ and the
width $\Gamma$ for the $^{6}$Be g.s. as a function of the basis size $%
K_{\max}$.}
\label{fig:en-con-k}
\end{figure}

The convergences of the predicted resonance energy and width as a function
of the hyperspherical basis size are shown in Fig.~\ref{fig:en-con-k} for
each of the potential sets. In all cases, our calculations are fully
converged. The resonance energies are forced to approach the experimental
value $E_{3r}=1.371$ MeV. This is achieved by fine tuning the
phenomenological potential of Eq.~\ref{eq:pot3}; this is a necessary
approach in order to provide reasonable predictions for the decay
characteristics. We can see that while the calculations with P1 and P2 (SBB
potential in the $\alpha $-$p$ channel) are in good agreement with each
other and with the experimental value, the width obtained with P3 (SBBM
potential) is far too large.

An expected feature observed here is the much slower convergence of the
calculations with a realistic potential in the $NN$ channel. An important,
but often disregarded fact, which one can see in Fig.~\ref{fig:en-con-k}, is
the much slower convergence of the width as compared to the energy. This
means that, \emph{in general, an energy convergence does not guarantee the
convergence of other important characteristics}. As we will see in Sect.~\ref%
{sec:com-modis}, the situation with momentum distributions is even more
complicated than it is for the widths.

\begin{figure}[tbp]
\includegraphics[width=0.244\textwidth]{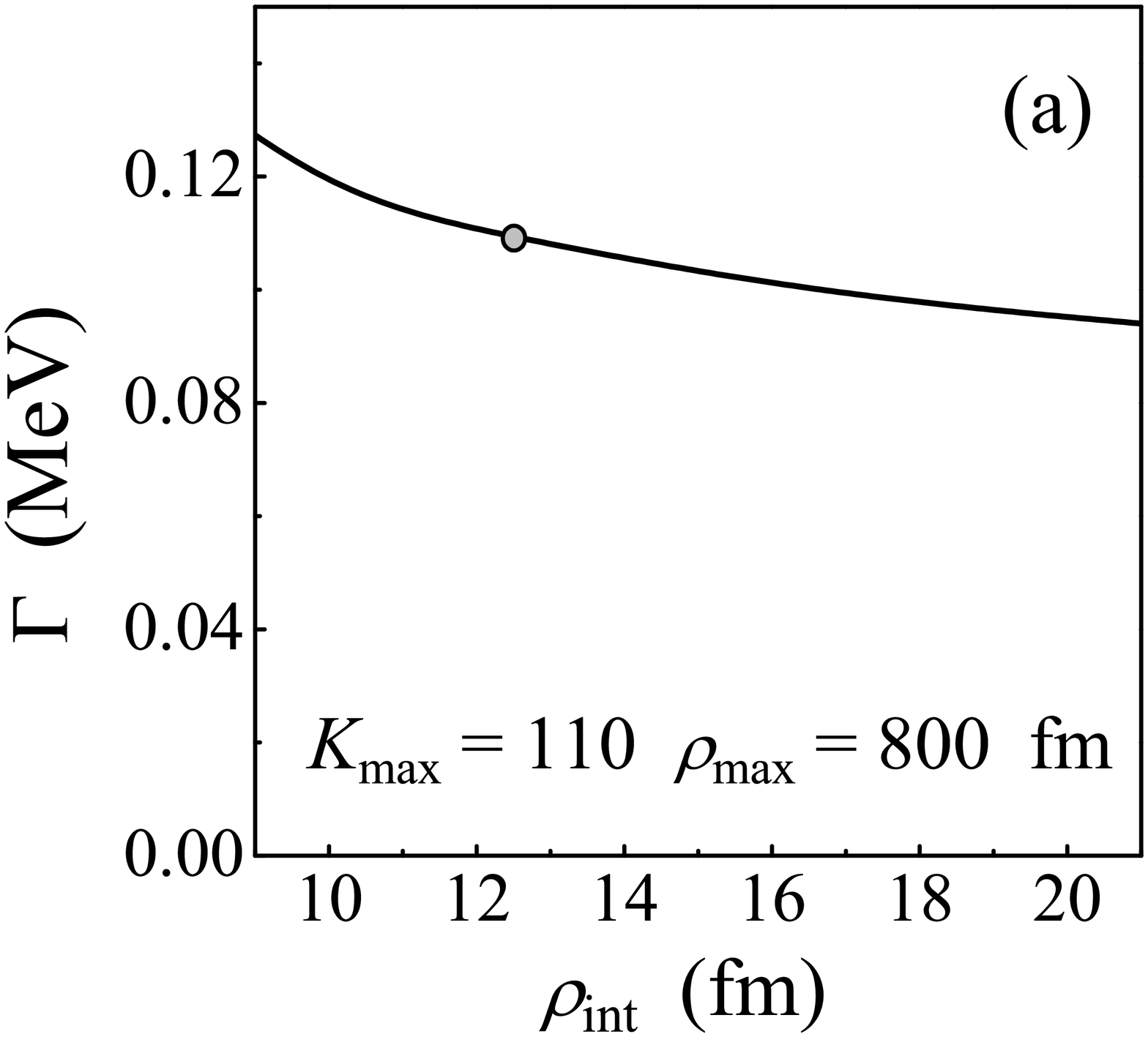} %
\includegraphics[width=0.23\textwidth]{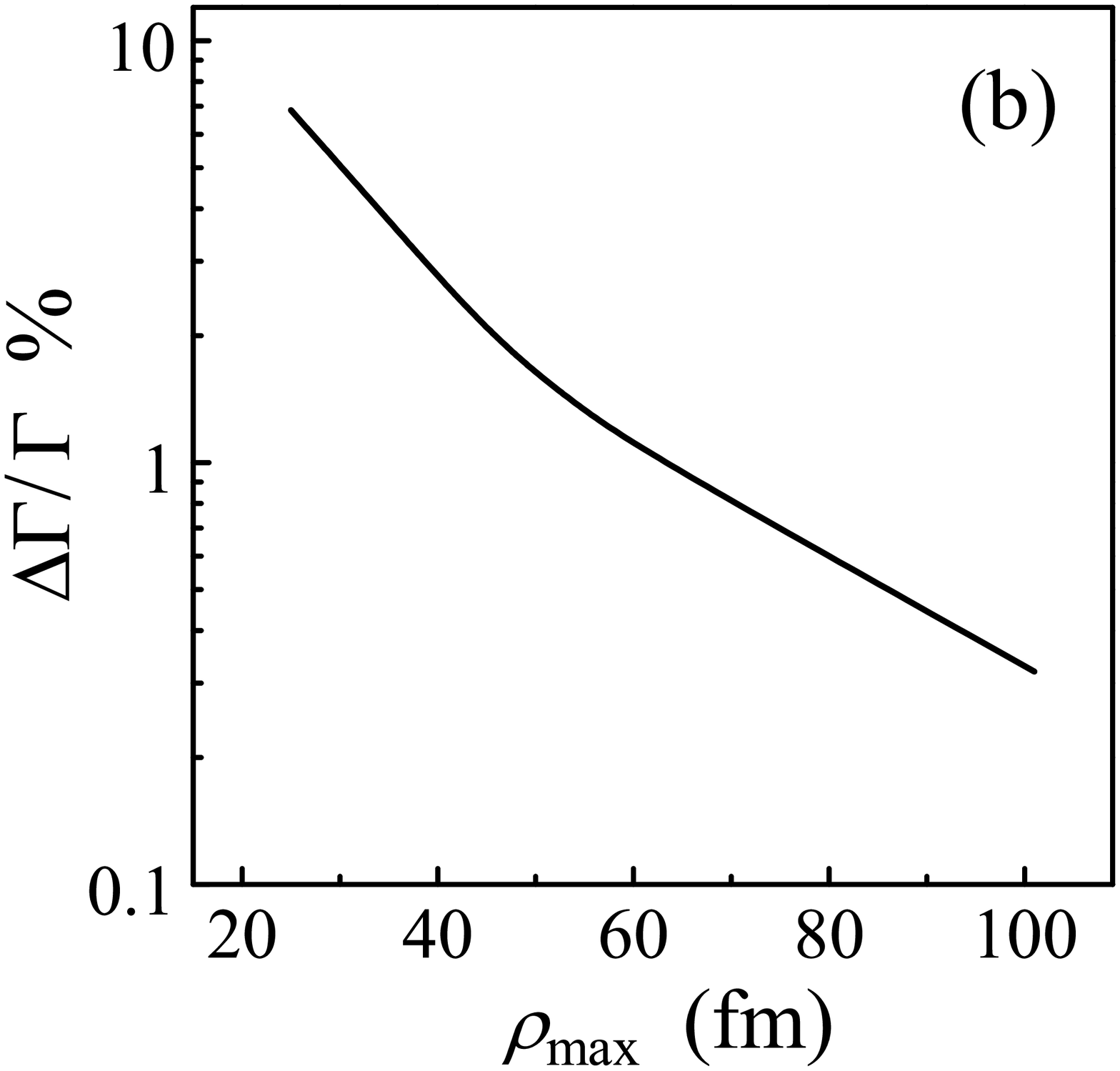}
\caption{(a) Sensitivity of the width as defined in Eq.~\protect\ref{eq:nat}
to the size of the ``internal region'' $\protect\rho_{\text{int}}$. The dot
shows the value of $\protect\rho_{\text{int}}$ at which this width coincides
with that defined via the cross-section profile Eqs.~\protect\ref{eq:ste-6li}%
, \protect\ref{eq:cross}, \protect\ref{eq:sig-prof}, and \protect\ref%
{eq:g-ot-e}. (b) Relative precision of the width as a function of the
matching radius $\protect\rho_{\max}$.}
\label{fig:rho-int-max}
\end{figure}

The sensitivity of the width to a number of the other parameters in the
calculations is demonstrated in Fig.~\ref{fig:rho-int-max}. Figure~\ref%
{fig:rho-int-max}(a) shows the sensitivity of the width defined by Eq.~\ref%
{eq:nat} to the size $\rho _{\text{int}}$ of the region where the internal
normalization is calculated. The stability of the calculations to the
dynamical range $\rho _{\max }$ is demonstrated in Fig.~\ref{fig:rho-int-max}%
(b). To attain $1\%$ numerical precision in the width calculations, we need
to go beyond 60~fm in the hyperradius $\rho $.

\subsection{Features of the momentum distributions in $^{6}$Be}

\label{sec:theory}


The correlations in the decay of $^{6}$Be include both the generic features
of the $2p$ decays, as discussed earlier in Refs.~\cite{gri03a,gri03c,gri07},
and some peculiarities which we present in more detail now. For nuclear
states with $J\leq 1/2$ (as is the case for $^{6}$Be g.s.\ decay), the
three-body correlations can be completely described by 2 parameters. There
are a total of 9 degrees of freedom for three particles in the final state.
Of these, three describe the center-of-mass motion, three describe the Euler
rotation of the decay plane (for $J\leq 1$ all its orientations are
quantum-mechanically identical), and the three-body decay energy is fixed.
Thus we are left with two parameters to describe the correlations. It is
convenient to choose the energy distribution parameter $\varepsilon $
between any two of the particles and the angle $\theta _{k}$ between the
Jacobi momenta:
\begin{equation}
\varepsilon =E_{x}/E_{T}\quad ,\quad \cos (\theta _{k})=\frac{\mathbf{k}%
_{x}\cdot \mathbf{k_{y}}}{k_{x}\,k_{y}}.  \label{eq:corel-param}
\end{equation}%
These parameters can be constructed in any Jacobi system and for $^{6}$Be
there are two ``irreducible'' Jacobi systems, called ``T'' and ``Y'' , see
Fig.\ \ref{fig:jacobi}. The distributions constructed in different Jacobi
systems are just different representations of the same physical picture.
However, different aspects of the correlations may be better revealed in a
particular Jacobi system.

\begin{figure}[tbp]
\includegraphics[width=0.49\textwidth]{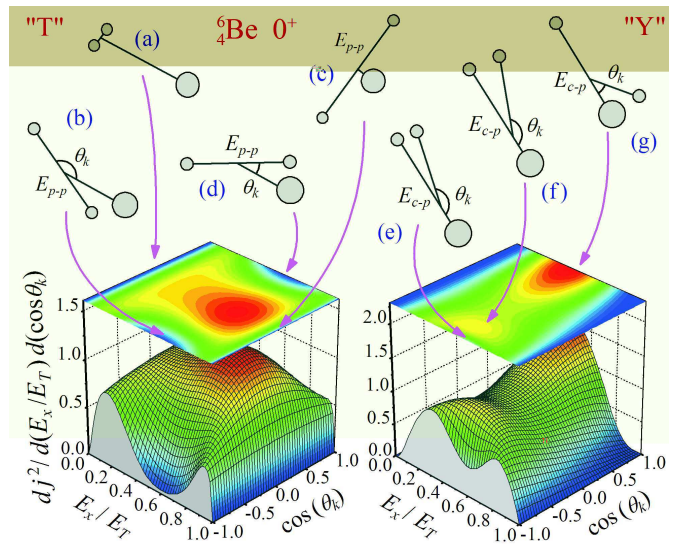}
\caption{(Color online) Complete correlation picture for $^{6}$Be g.s.\
decay, presented in ``T'' and ``Y'' Jacobi systems.}
\label{fig:corel-all}
\end{figure}

Predictions for the complete correlation picture of $^{6}$Be g.s. decay are
shown in Fig.~\ref{fig:corel-all} for both the ``T'' and ``Y'' Jacobi
systems. Schematic figures are included in this figure to help in
visualizing the correlations associated with different regions of the Jacobi
plots. The main features of these distributions are:

\begin{enumerate}
\item The energy distribution in the ``T'' system has a double-humped profile
which is an indication of the $[p^{2}]$ configuration dominance which was
pointed out in very early papers on $^{6}$Be \cite{boc87,dan87,boc89}. This
double-humped configuration is expressed more in coordinate space (see
the internal region in Fig.~\ref{fig:corel-dens}) and only marginally
``survives'' in the asymptotic region. The internal peaks in Fig.~\ref%
{fig:corel-dens} have the special names of ``diproton'' (protons are close
to each other) and ``cigar'' (protons are in-line with $\alpha $-particle)
configurations \cite{dan91}.

\item There are kinematical regions where the presence of particles is
suppressed due to Coulomb repulsions. Strong suppression in the $\alpha $-$p$
channel in regions (b) and (d) and a smaller suppression in the $p$-$p$
channel in region (e) are predicted.

\item There are enhancements due to the $p$-$p$ final-state interaction in
regions (a) and (f). The $^{5}$Li g.s. resonance in the $\alpha $-$p$
channel is not accessible for decay. However, some hint of its presence can
be obtained from the enhancement in region (g). This is a ``back-to-back''
configuration, where protons fly in the opposite directions. However, the
reason for the enhancement of such a configuration is not fully understood.

\item The angular dependence in the ``T'' system almost vanishes for regions
(a) and (c) ($E_{x}/E_{T}\sim 0$ and $E_{x}/E_{T}\sim 1$). It is clear that
in the limit $E_{x}/E_{T}\rightarrow 0$ and $E_{x}/E_{T}\rightarrow 1$ the
dependence on the relative orientation of $\mathbf{k}_{x}$ and $\mathbf{k}%
_{y}$ should become degenerate. However at intermediate values of $%
E_{x}/E_{T}$, this dependence is very pronounced.

\item The total-energy distribution in the ``Y'' system (see Fig.~\ref%
{fig:corel-dep-pot} for the projected distributions) is almost a symmetric
bell-shape. This is the energy distribution between the core and one of the
protons and its symmetry reflects the symmetry between protons. In
heavy two-proton emitters, this distribution becomes very narrow and almost
completely symmetric.
\end{enumerate}

\begin{figure}[tbp]
\includegraphics[width=0.34\textwidth]{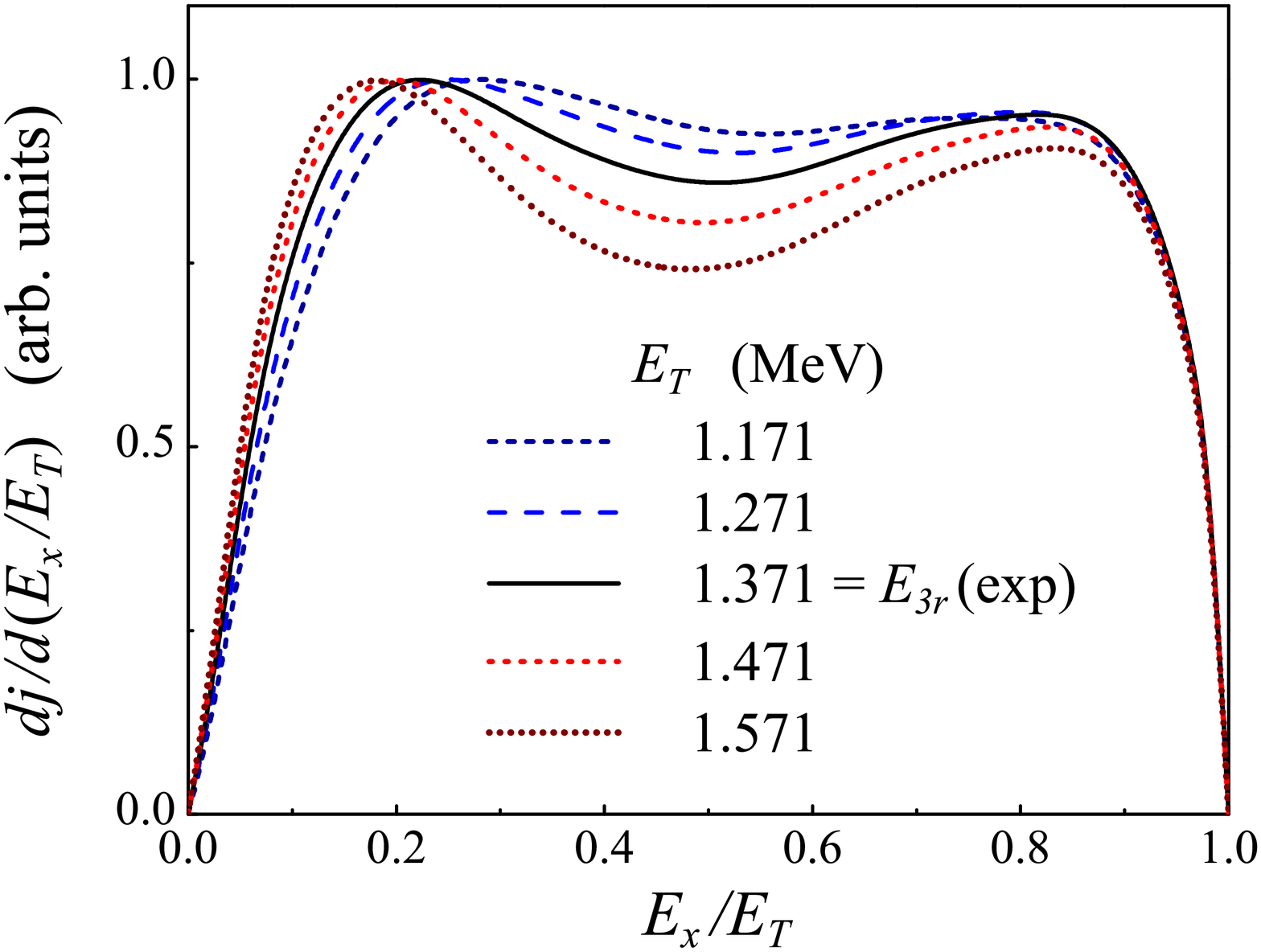}
\caption{(Color online) Dependence of energy distribution between the proton
(``T'' system) in the decay of $^{6}$Be g.s. on the decay energy $E_T$.}
\label{fig:corel-en-dep}
\end{figure}

The correlation predictions shown in Fig.~\ref{fig:corel-all} are obtained
on resonance. The dependence the of energy correlation on the decay energy
of $^{6}$Be is demonstrated in Fig.~\ref{fig:corel-en-dep}. The
double-humped shape of this spectrum becomes less pronounced when the energy
decreases. With smaller energy, the relative contribution of the $[s^{2}]$
configuration to the decay grows compared to the $[p^{2}]$ configuration.
The latter has an additional centrifugal component to the barrier and its
contribution to the width should be suppressed at low energies. The pure $%
[s^{2}]$ configuration should produce a featureless \textquotedblleft
phase-volume\textquotedblright\ energy distribution
\begin{equation}
dj/dE_{x}\sim \sqrt{E_{x}(E_{T}-E_{x})}\;.
\end{equation}

\begin{figure}[tbp]
\includegraphics[width=0.34\textwidth]{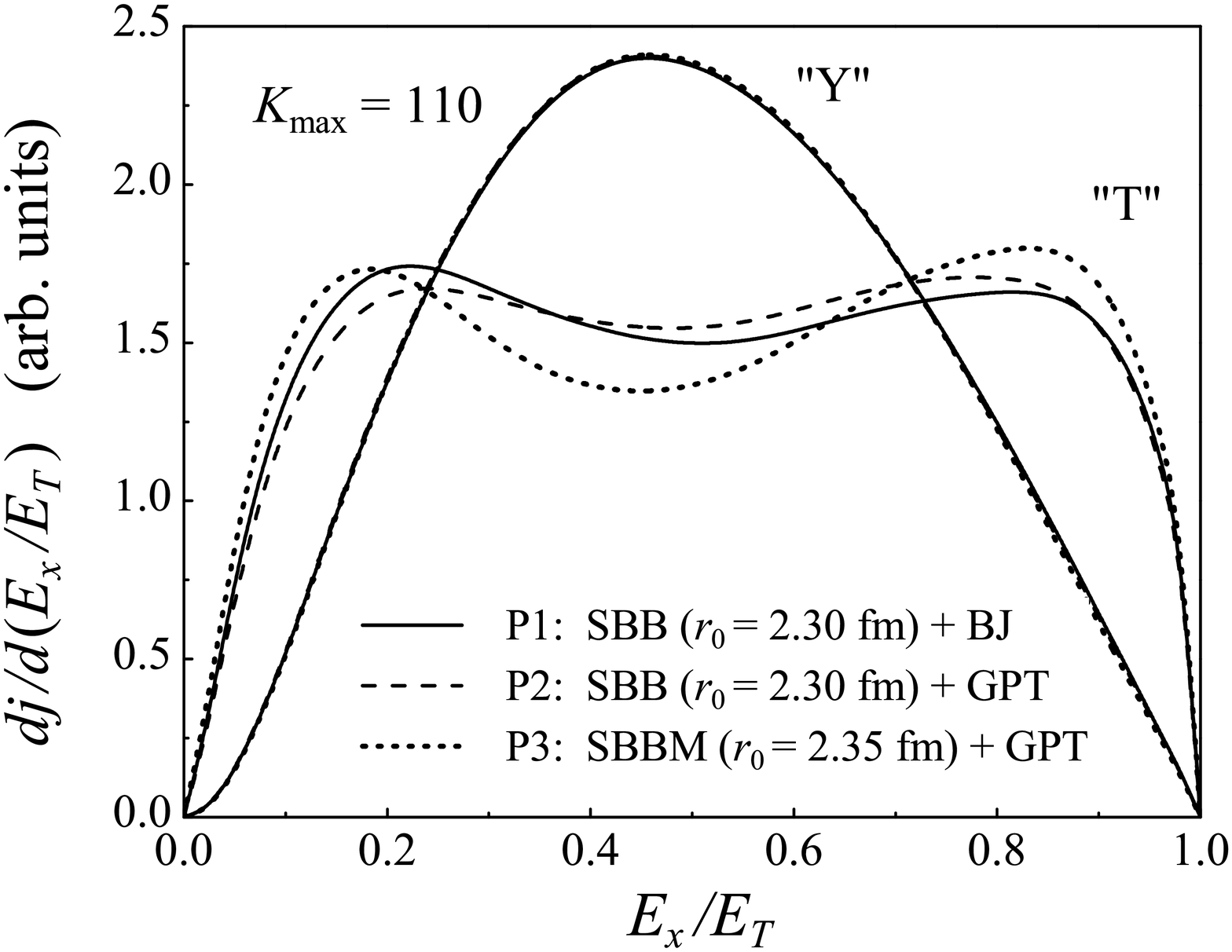} %
\includegraphics[width=0.34\textwidth]{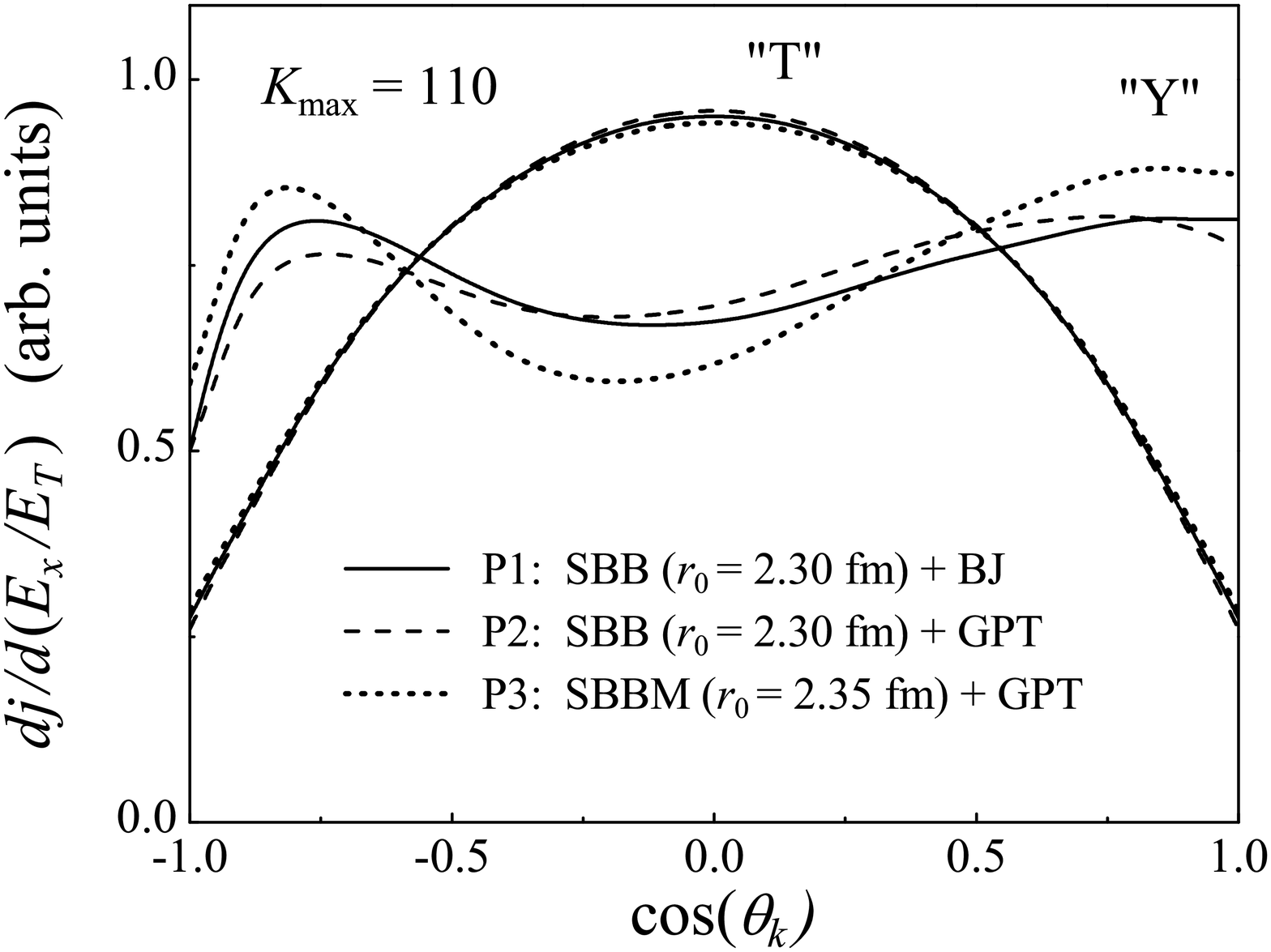}
\caption{Sensitivity of the energy and angular distributions in the decay of
$^{6}$Be g.s.\ to the choice of the potential set. Results are shown for
both the ``T'' and ``Y'' Jacobi systems.}
\label{fig:corel-dep-pot}
\end{figure}

The sensitivity of the projected distributions to the choice of the
potential set P1-P3 is demonstrated in Fig.~\ref{fig:corel-dep-pot}. The
angular distribution in the ``T'' system and the energy distribution in
``Y'' systems are practically insensitive to this choice. The other
projected distributions demonstrate sensitivity on the level of $10-15\%$.
However, local differences in certain kinematical regions are much larger.

Figures~\ref{fig:corel-all} and \ref{fig:corel-dep-pot} demonstrate what we
call the ``softness'' of the $^{6}$Be system: minor variations in the
conditions or computational details lead to a noticeable variations in the
observable properties. Heavier $2p$ emitters appear to be much ``stiffer''
in this respect.


\subsection{Convergence of the momentum distributions}

\label{sec:com-modis}

\begin{figure}[tbp]
\includegraphics[width=0.34\textwidth]{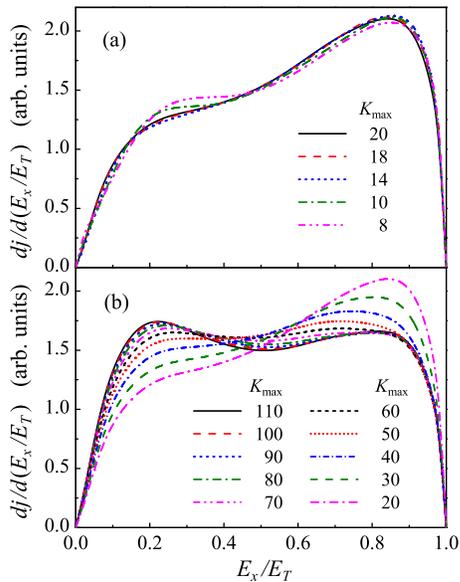}
\caption{(Color online) Convergency of the ``T'' energy distribution in the
decay of $^{6}$Be g.s. as a function of the basis size $K_{\max}$.}
\label{fig:corel-con-k}
\end{figure}
\begin{figure}[tbp]
\includegraphics[width=0.34\textwidth]{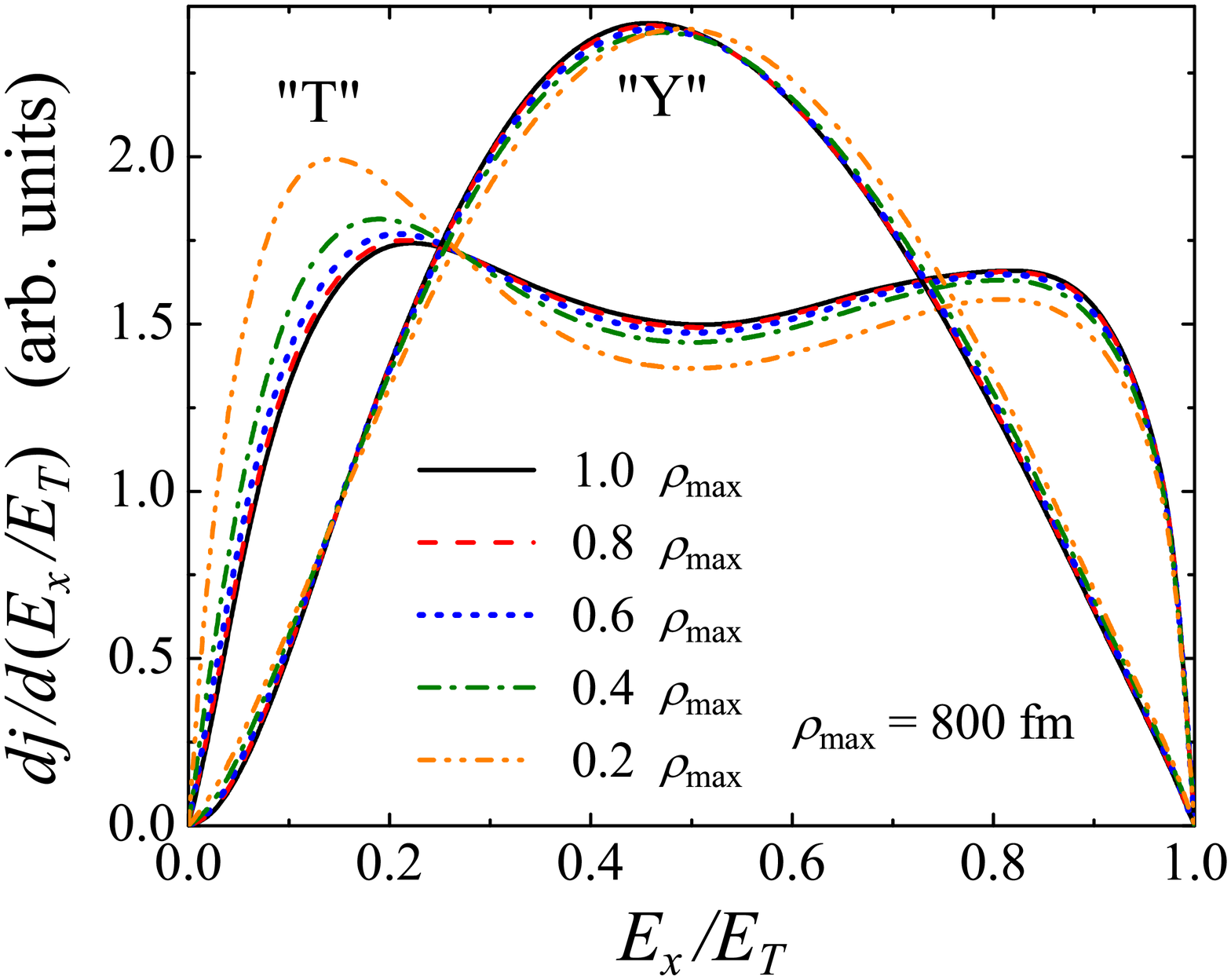}
\caption{(Color online) Dependence of energy distribution in the decay of $%
^{6}$Be g.s.\ on the maximal dynamic range of the calculation $\protect\rho%
_{\max}$. For the ``T'' Jacobi system, $E_x$ is energy between two protons
and in the ``Y'' Jacobi system, $E_x$ is energy between core and one of the
protons.}
\label{fig:corel-rho-dep}
\end{figure}

In our calculations there are two projected distributions which are practically 
insensitive to convergence issues (the angular distribution in the ``T'' system 
and the energy distribution in the ``Y'' system). The other two distributions 
(the angular distribution in the ``Y'' system and the energy distribution in the 
``T'' system) demonstrate strong sensitivity. The convergence of the energy 
distributions are illustrated in Figs.~\ref{fig:corel-con-k} and 
\ref{fig:corel-rho-dep}.

The convergence of the energy distribution between protons has a very curious 
character, see Fig.~\ref{fig:corel-con-k}. From $K_{\max }=8$ to $K_{\max }=22$ 
this distribution is very stable [several curves almost coincide, see 
Fig.~\ref{fig:corel-en-dep}(a)]. Then from $K_{\max }=24$ to $K_{\max }\sim 70$ 
the distribution changes qualitatively, and up to $K_{\max}\sim 100$ there is 
still a noticeable variation [Fig.~\ref{fig:corel-en-dep}(b)]. Hopefully with 
$K_{\max }=110$, we have a well converged distribution. Calculations with small 
basis sizes (e.g., $K_{\max }\leq 70$) for $^{6}$Be should provide a 
qualitatively wrong energy distribution in the ``T'' system. Similarly for the  
angular distribution in the ``Y'' system.

This ``softness'' of the $^{6}$Be system makes it a very complicated object to 
study. Minimum basis sizes which provide convergence for the energy and width 
are far from sufficient for calculations of momentum distributions. This is a 
feature which we probably do not face in heavier $2p$ emitters as the Coulomb 
interaction in the core-$p$ channel plays a more dominant role in the decay 
dynamics.

The radial convergence of the energy distributions is illustrated in 
Fig.~\ref{fig:corel-rho-dep}. Calculations with $\rho _{\max }<300$ fm are 
clearly insufficient to stabilize the distribution. However by $\rho 
_{\max}=800$ fm, the distributions seem to be well converged. Could there be 
some noticeable modifications of the distributions due to further propagation in 
the long-range Coulomb field? This question was analyzed in Ref.~\cite{gri03c} 
for $^{45}$Fe using the classical trajectory approach. The complete 
stabilization takes place in $^{45}$Fe at $\rho \sim (3-6)\times 10^{4}$ fm, 
with a major part of the effect originating at $\rho \lesssim 1\times 10^{4}$ 
fm. The decay energies of $^{6}$Be and $^{45}$Fe g.s.\ are similar and the 
Coulomb interaction is $\sim 12$ times weaker in $^{6}$Be. Therefore, the 
majority of the long-range effects should be taken into account in calculations 
with $\rho _{\max }\sim 1000$~fm. The $^{6}$Be calculations of this work were 
typically done with $\rho _{\max }=1200$ fm.


\subsection{Structure of the $^{6}$He and $^{6}$Be g.s.}


From another point of view, one can benefit from the ``softness'' of $^{6}$Be 
system. The high sensitivity of the observables to the details of the model 
ingredients increase our ability to discriminate these features and hence 
improve our ability to elucidate the details of the nuclear structure.

\begin{figure*}[tbp]
\includegraphics[width=0.344\textwidth]{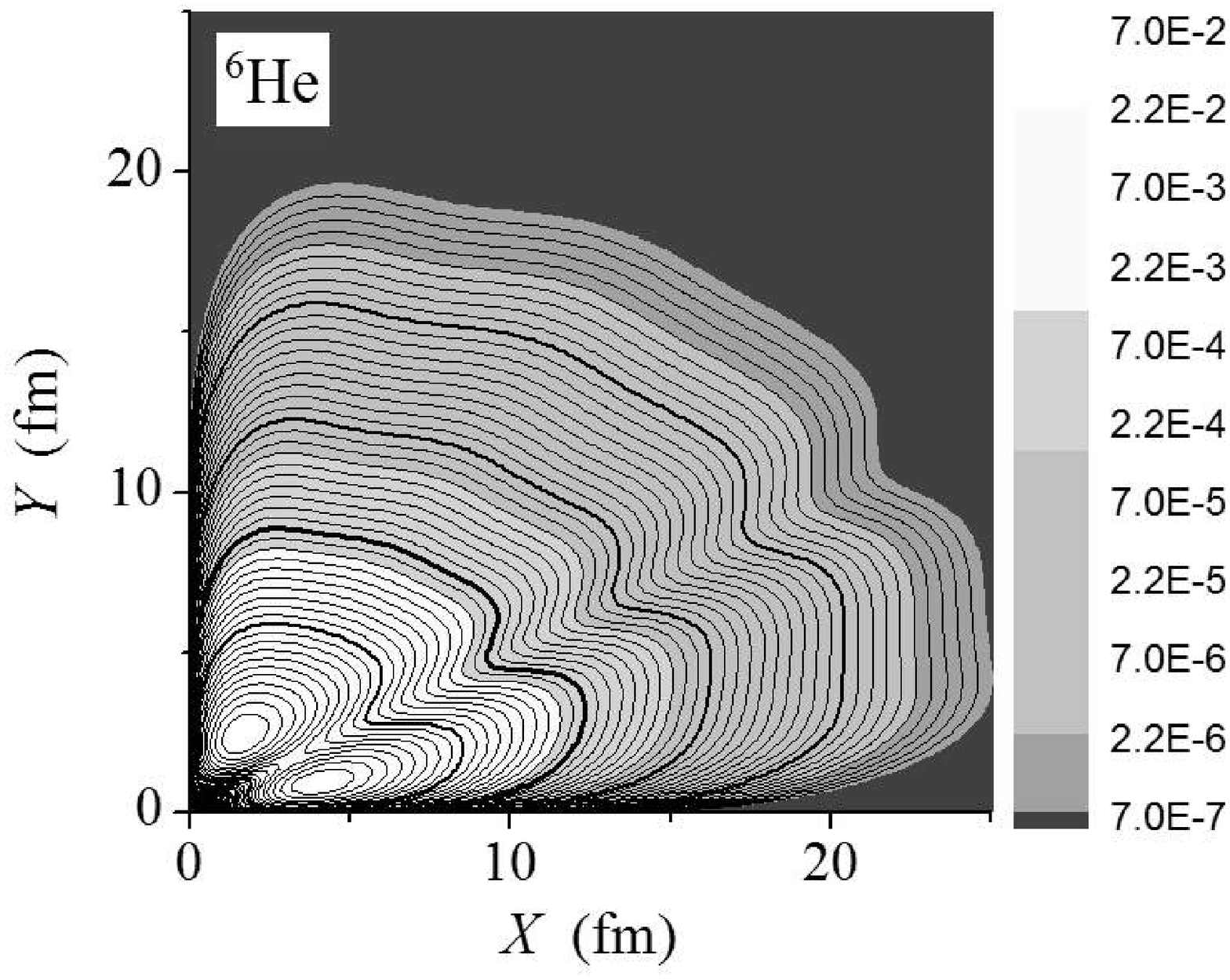} 
\includegraphics[width=0.317\textwidth]{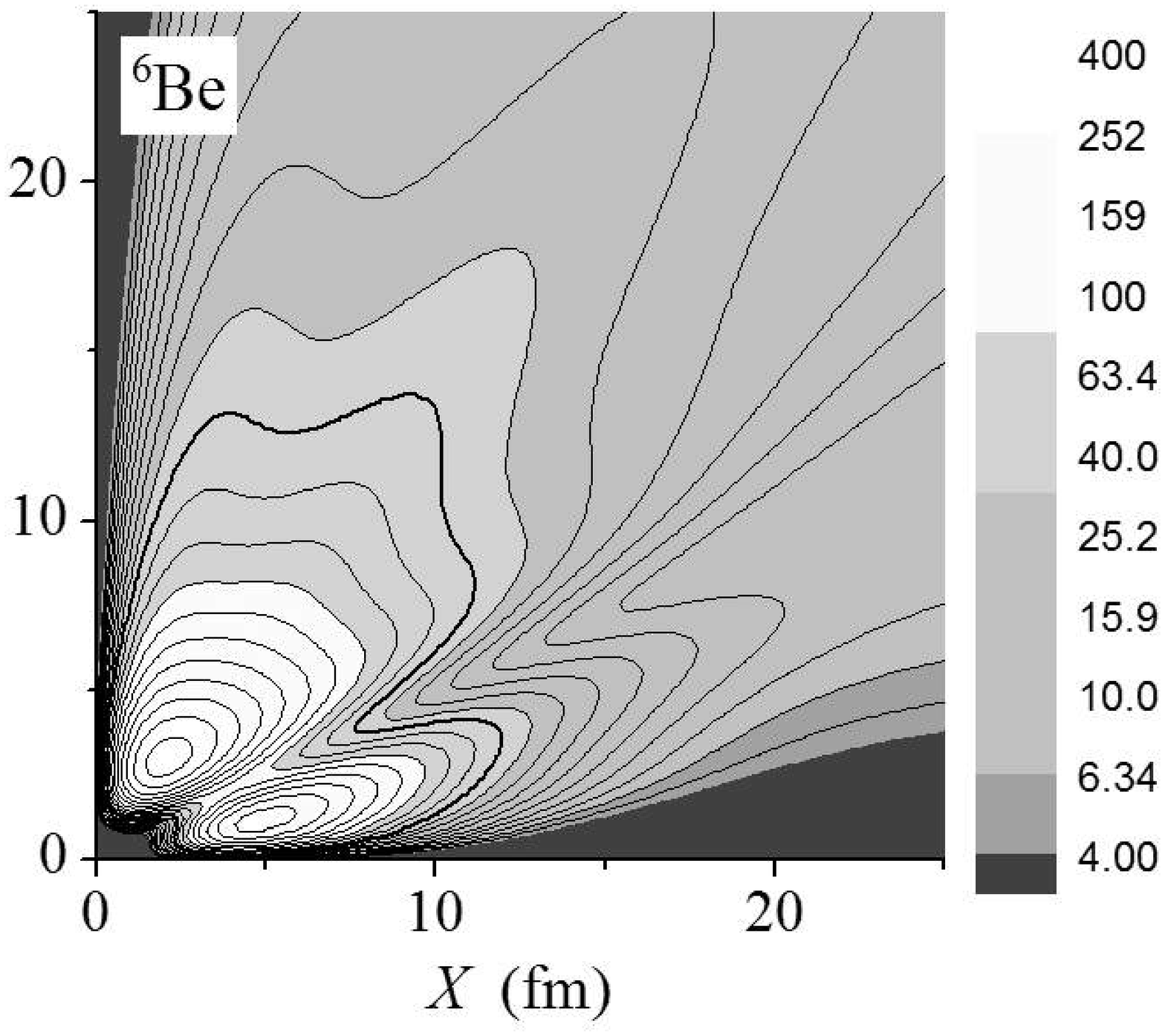} 
\includegraphics[width=0.32\textwidth]{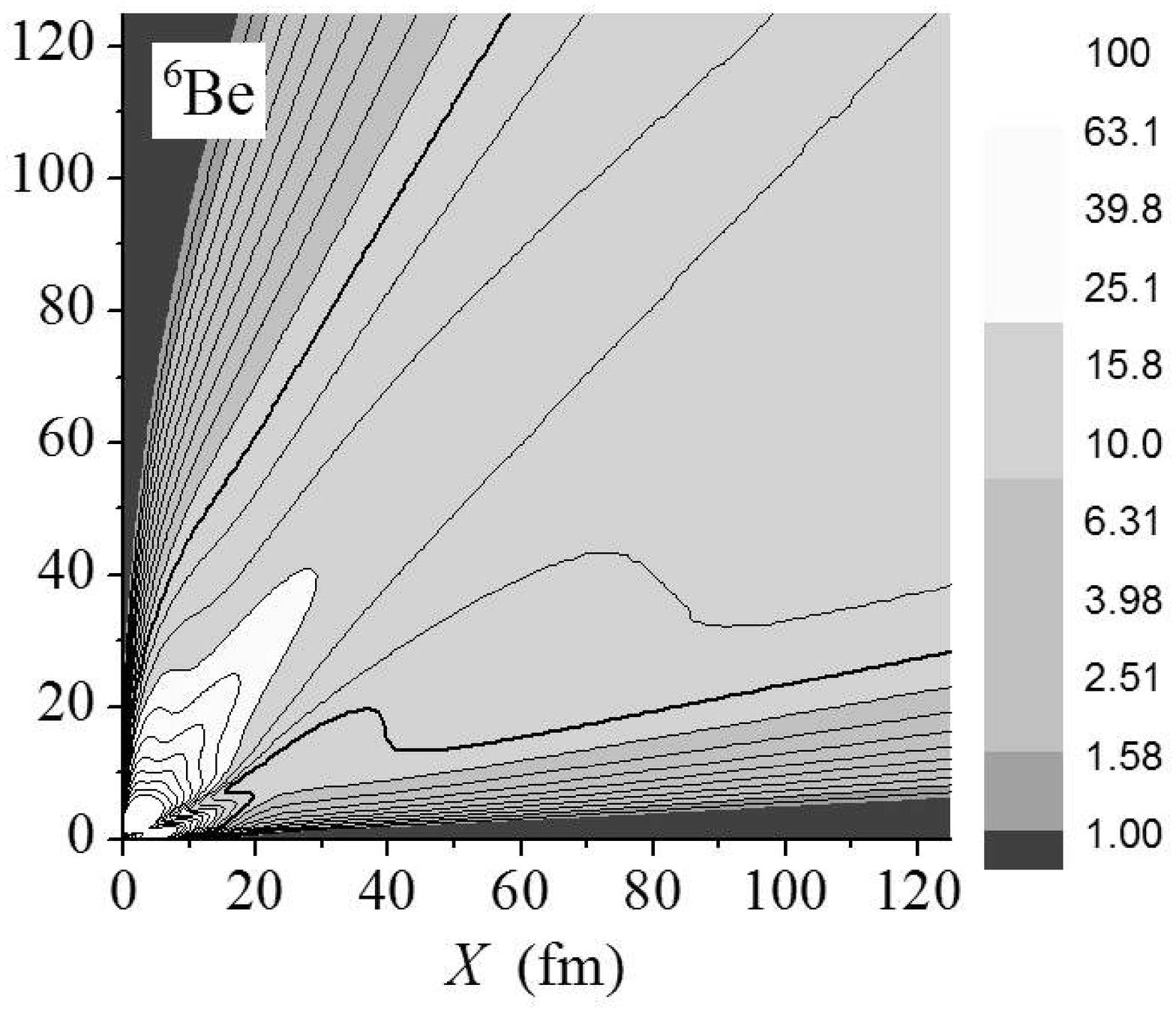}
\caption{Spatial correlation densities $|\Psi(X,Y)|^2$ for the $^{6}$He and $%
^{6}$Be g.s.\ WFs in the ``T'' system. For $^{6}$Be, two panels provide the
view in different radial ranges. Pay atention to difference in the scales.
The variation from the top to the thick contour line in the $^{6}$He panel
corresponds to the whole scale variation in the two $^{6}$Be panels. }
\label{fig:corel-dens}
\end{figure*}

\begin{table}[b]
\caption{Weights of the shell-model-like configurations $[l^2]$ in the $^{6}$He 
and $^{6}$Be g.s.\ WFs in percent for the Jacobi ``Y'' system. The
normalizations of the $^{6}$Be components are found for integration radius 
$\protect\rho_{\text{int}}=12.5$ fm.}
\label{tab:struc-y}
\begin{ruledtabular}
\begin{tabular}[c]{lcccccc}
  &   \multicolumn{3}{c}{$^6$He} &  \multicolumn{3}{c}{$^6$Be}  \\
$[l^2]$ & P1 & P2 & P3 & P1 & P2 & P3  \\
\hline
$[s^2]$ & 8.11  & 8.58 & 8.35  & 10.54 & 11.15  & 10.84  \\
$[p^2]$ & 90.91 & 90.30 & 90.37 & 87.98 & 87.18  & 87.17  \\
$[d^2]$ & 0.47  & 0.53 & 0.61  & 0.69  &  0.77  & 0.95  \\
$[f^2]$ & 0.41  & 0.43 & 0.50  & 0.60  &  0.65  & 0.77  \\
\end{tabular}
\end{ruledtabular}
\end{table}

Detailed information about the $^{6}$He and $^{6}$Be g.s.\ WFs is provided in 
Table \ref{tab:struc-t}. In general, there is high degree of isobaric symmetry 
between the $^{6}$He and $^{6}$Be WFs in the internal region. This is not true, 
however, for the $K=0$ component, which differs the most. The reason for this is 
shown in Fig.~\ref{fig:wfs} where the magnitude of the $K=0$ WF in asymptotic 
region is comparable to its magnitude in the nuclear interior. Hence the nuclear 
boundary is not defined for this component in $^6$Be. This is also seen in Table 
\ref{tab:struc-y}, which provides the information about the WF in approximate 
``shell model'' terms. After looking at the radial behavior of the WF's 
components in Fig.~\ref{fig:wfs}, we find that the concept of isobaric symmetry 
is relevant here strictly speaking \emph{only} for the most interior region of 
the WF ($\rho <4-5$ fm). Beyond this point the radial behavior in $^{6}$He and 
$^{6}$Be differ drastically.

The weights of the components in Tables \ref{tab:struc-t} and \ref{tab:struc-y} 
are in \emph{very} good relative agreement for the different
potential sets P1-P3. Evidently these major features of the structure are
not that sensitive to the fine details of the interactions.

\begin{table}[b]
\caption{Radial properties of the $^{6}$He g.s.\ WF and some observables
obtained for $^{6}$He and $^{6}$Be g.s.\ with potentials P1, P2, P3.}
\label{tab:observ}
\begin{ruledtabular}
\begin{tabular}[c]{ccccc}
value & P1 & P2 & P3 & Exp.\   \\
\hline
$\langle \rho \rangle $  (fm)  & 5.088 & 5.156 & 5.491  &    \\
$\langle r_{NN} \rangle$  (fm) & 4.482 & 4.502 & 4.884  &    \\
$\langle r_{cN} \rangle$ (fm)  & 4.113 & 4.172 & 4.430  &    \\
$\langle r_{N} \rangle$ (fm)   & 3.211 & 3.248 & 3.469  &    \\
$\langle r_{c} \rangle$ (fm)   & 1.321 & 1.171 & 1.232  &    \\
$r_{\text{mat}}$    (fm)       & 2.396 & 2.421 & 2.540  & $2.30\pm 0.07$
\cite{ege01} \\
             &  &  &   & $2.48\pm 0.03$ \cite{oza01} \\
$r_{\text{ch}}$\footnotemark[1] (fm)  & 2.103 & 2.012 & 2.048 & $2.054\pm 0.014$
\cite{wan04} \\
$r_{\text{ch}}$\footnotemark[2] (fm)  & 2.113 & 2.043 & 2.079 &  $2.068\pm
0.011$ \cite{mue07} \\
$B_{GT}(^6$He$\rightarrow^6$Li) & 5.004  & 5.058 & 4.930 & $4.745 \pm 0.009$
\cite{til02} \\
$\Delta E_{\text{coul}}$ (MeV) & 2.351 & 2.302 & 2.111  &  2.344 \cite{til02} \\
$\Gamma (^6$Be$_{\text{g.s.}})$ (keV) & 98 & 112 & 154 & $92\pm 6$
\cite{til02}\\
\end{tabular}
\end{ruledtabular}
\footnotetext[1]{Theoretical
values in this row are obtained using the generally accepted value for the
neutron charge radius $r^2_{\text{ch}}(n)= - 0.1161$ fm$^2$.}
\footnotetext[2]{Theoretical
values in this row are obtained using $r^2_{\text{ch}}(n)= 0.012$ fm$^2$
\cite{mue07}.}
\end{table}

It can be seen that the partial widths $\Gamma _{i}$ of the $^{6}$Be WF
components in Table \ref{tab:struc-t} are drastically different as compared
to weights $N_{i}$ in the internal region. This is a reflection of
complicated dynamics in $2p$ decays, the WFs are strongly ``rearranged'' in
the subbarrier region and by the long-range Coulomb pairwise
fields. The $^{6}$He and $^{6}$Be WF correlation densities are shown in Fig.~%
\ref{fig:corel-dens}. The WFs are nearly identical in the internal region,
while in the asymptotic region for $^{6}$Be we can clearly see how this
``rearrangement'' is taking place. Comparing different potential sets P1-P3
in Table \ref{tab:struc-t}, we see that P1 and P2 calculations are almost
identical, while the major partial width in P3 differs strongly. We conclude
that the decay dynamics is mainly defined by core-$p$ interaction.

Geometric properties of the $^{6}$He g.s.\ WF and several observables obtained 
for $^{6}$He and $^{6}$Be g.s.\ are shown in Table \ref{tab:observ}. The root 
mean square values are given for $\rho$, $r_{NN}$ (distance between valence 
nucleons), $r_{cN}$  (distance between nucleon and core), $r_{N}$ (distance 
between valence nucleon and c.m.), $r_{cN}$  (distance between core and c.m.).  
The differences between these geometric characteristics for P1 and P2 are 
typically around $1\%$. In the case of P3, the differences are significantly 
larger. The $B_{GT}$ values obtained with P1-P3 also agree within $1.5\%$, but 
all differ more from the experimental value. Here, the ``experimental'' $B_{GT}$ 
value is obtained using the $^{6}$He lifetime $\tau _{1/2}=806.7\pm 1.5$ ms 
\cite{til02}, and the $\beta $-decay constants of $ft(0^{+}\rightarrow 
0^{+})=3072.40$~s and $\lambda =1.268$. It has already been discussed in the 
literature that the $4-7\%$ disagreement here could be connected with both the 
WF quality and the renormalization of the weak constant \cite{dan91}. Therefore, 
we give no definite conclusion about quality of the models here.

The next most precisely known characteristic for $^{6}$He is its charge radius. 
Recent studies have defined $r_{\text{ch}}$ with increasing precision 
\cite{wan04,mue07}. The relative uncertainty of this value is now about $0.5\%$, 
while variations in the calculated value are around $4\%$ for P1-P3. However, 
comparison of this value with those theoretically calculated is not completely 
model independent. The theoretically calculated charge radius of $^{6}$He is 
noticeably sensitive to the neutron charge radius. The latter is inferred 
theoretically, rather than measured experimentally. This means that there exists 
considerable systematic uncertainty in the determination of the charge radii. 
According to our estimates, this uncertainty can be as large as $2\%$. This fact 
somewhat relieves the constrains on the WF connected with this observable. One 
can see in Table \ref{tab:observ} that the P2 and P3 calculations, containing 
realistic $NN$ potentials can be regarded as consistent with the experiment.

The matter radius of $^{6}$He is defined in the cluster model using the
matter radius of the $\alpha $-particle. The value $r_{\text{mat}}(\alpha
)=1.464$ fm is derived from the charge radius $r_{\text{ch}}(\alpha )=1.671$
using the neutron and proton charge radii; $r_{\text{ch}}^{2}(n)=-0.1161$ 
fm$^{2}$, $r_{\text{ch}}(p)=0.875$ fm. The experimental data on matter radii
have large a systematic uncertainty. This is probably the reason for the
controversial signal obtained in different experiments (see two examples in
Table \ref{tab:observ}). This observable so far does not seem to have
discriminative power for theoretical models.

The Coulomb shift $\Delta E_{\text{coul}}$ and $^{6}$Be g.s.\ width obtained 
with P1 and P2 are in a good agreement with experiment. Some overestimation of 
the width in the three-body cluster model can be expected due to the admixture 
of different configurations in $^{6}$Be WF. The weight of such admixtures can be 
estimated as $6-14\%$, based on the P1 and P2 widths. However, the Coulomb shift 
and width obtained with P3 are clearly not acceptable. Our overall feeling is 
that the cumulative information on $^{6}$He and $^{6}$Be g.s.\ is sufficient to 
choose P2 as the only acceptable potential.


\section{Theoretical discussion}


As we have already mentioned, most of the attention in the studies of the $A$%
=6 isobar has been paid to $^{6}$He. Even in the studies of $^{6}$Be, there
are only few works which studied it's width. In addition, there has been are
only limited studies of the $^{6}$Be g.s.\ decay correlations. The first
consistent calculations of the $^{6}$Be three-body decay width were
performed in Ref.~\cite{dan93} using the integral formalism. In papers 
\cite{gri00b,gri01}, the quantum-mechanical formalism for two-proton
radioactivity and Coulombic three-body decay studies was developed. In these
papers, the integral formalism was criticized in application to the decays
of systems with strong three-body Coulomb interactions and a more
preferable way to calculate widths was proposed [see, Eq.~\ref{eq:nat}]. The
value $\Gamma =90$ keV was obtained in Ref.~\cite{gri00b} with the P1
potential ($K_{\max }=20$), which as we can see in Fig.~\ref{fig:en-con-k},
is reasonably well converged.

In our approach, the effects of antisymmetrization are taken into account in
a simplified way. However, there are studies that treated the $^{6}$Be decay
as a 6-body problem. In RGM calculations \cite{cso94}, the $^{6}$Be width of
$\Gamma =160$~keV for $E_{3r}=1.52$~MeV was found using the complex scaling
method. Scaling this value to the experimental $2p$ decay energy with the
help of Fig.~\ref{fig:pen-ot-e} we obtain $\Gamma =125$~keV which is
considerably larger than the experimental value. An interesting algebraic
method was developed for studies of $^{6}$Be decay in Ref.~\cite{vas01}.
Here, the hyperspherical decomposition is used for the WF both in the
internal region (6-body HHs) and in the asymptotic region (three-body
cluster HHs). A calculated width of $\Gamma =72$ keV was obtained for 
$E_{3r}=1.172$ MeV which scales to $\Gamma =110$ keV at the experimental $2p$
decay energy. In addition, we can expect a $10-15\%$ reduction due to the
absence of the $S=1$ component in these calculations. This component is
important in the internal region, but does not contribute to the width
significantly. In addition, we can also expect roughly a factor of 2
increase due to the small basis size ($K_{\max }=10$) used in the asymptotic
region in Ref.~\cite{vas01}. According to Fig.~\ref{fig:en-con-k}, with $K_{\max 
}=10$ we can expect only $60\%$ of the width, at most. It seem that
Ref.~\cite{vas01} is more a concept demonstration, rather than a realistic
calculation. Therefore at the present moment, it is not possible to draw any
conclusions about importance of the 6-body effects in calculations of the 
$^{6}$Be decay properties.

The width of the $^{6}$Be g.s.\ was calculated in Ref.~\cite{des06} via a
method analogous to ours (hyperspherical harmonics), but having certain
technical differences. An approximate treatment of the $3\rightarrow 3$
scattering is introduced in this work and the width is extracted from the
energy behavior of the phase shifts. The width obtained was $\Gamma =65$~keV
for $E_{3r}=1.26 $ MeV which scales to $\Gamma =84$ keV at the experimental $2p$ 
decay energy. It can be found in Ref.~\cite{des06} that the calculation
does not seem converged. If we extrapolate from $\Gamma =84~$keV using the
convergence curves for P2, P3, then the value $\Gamma =110$ keV is obtained,
which is is a good agreement with our P2 result.

An important result of the present work is the clear demonstration that any
approach purporting to give satisfactory description of the $^{6}$Be g.s.\
decay properties should have a certain ``dynamic range'' both in radial and
functional spaces (see Table \ref{tab:range}). It can be found that not all
of these conditions are satisfied in these other works dedicated to $^{6}$Be.

Our calculations demonstrate a noticeable sensitivity of the observables in
the decay of $^{6}$Be g.s.\ to the ingredients of the model. Table 
\ref{tab:observ} demonstrates that this sensitivity is enhanced in $^{6}$Be
compared to $^{6}$He. Typical variations of the observables for $^{6}$He are 
$0.5-4\%$, while in $^{6}$Be there is about a $60\%$ difference in between
the widths calculated with P1 and P3. The tunneling process can be seen as a
kind of a ``quantum amplifier'', which drastically emphasizes minor features
in the structure. For that reason, it is possible that the indirect probe of
$^{6}$Be decay is a more sensitive tool for determining the halo properties
of $^{6}$He than direct investigations of $^{6}$He itself. We are referring
to precision measurements of the correlations in $^{6}$Be decay which are
discriminative with respect to the fine details of the momentum
distributions. In the experimental studies presented in this work, the
quality of the data is approaching fulfilment of such a high precision
request.

\begin{table}[b]
\caption{Minimal dynamical ranges of calculations required to provide
reasonably converged different observables for $^{6}$Be. Different basis
sizes are required for simplistic BJ and realistic GPT potentials in the $p$-$p$ 
channel.}
\label{tab:range}
\begin{ruledtabular}
\begin{tabular}[c]{cccc}
value  & $E_{3r}$ & $\Gamma$ & distributions \\
\hline
$\rho_{\max}$ (fm) & 20  & 60 & 300 \\
$K_{\max}$ (SBB+BJ)  & 16  & 30  & 80 \\
$K_{\max}$ (SBB+GPT) & 40  & 70  & 110 \\
\end{tabular}
\end{ruledtabular}
\end{table}


\section{Existing experimental knowledge about $^{6}$Be}


Very precise results about the energy and width of the $^{6}$Be g.s.\ were
obtained in the early studies: $E_{T}=1371(5)$ keV, $\Gamma=89(6)$ keV 
\cite{wha66}. The current value of the width is only slightly
different $\Gamma =92(6)$ keV \cite{til02}.

\begin{figure}[tbp]
\includegraphics[width=0.37\textwidth]{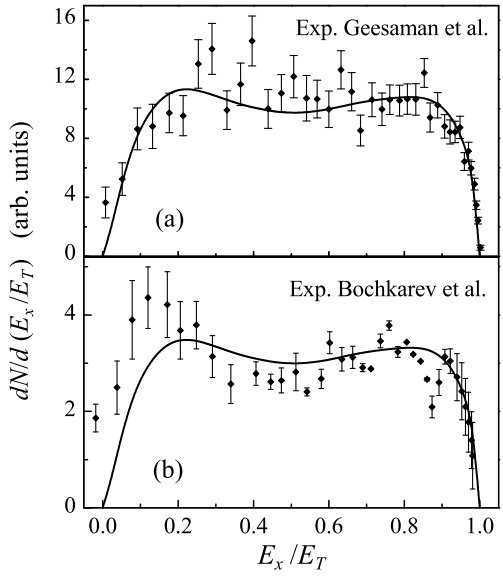}
\caption{Experimental energy distributions between protons in the decay of 
$^{6}$Be measured in (a) Ref.~\protect\cite{gee77} and (b) 
Ref.~\protect\cite{boc89}. The theoretical prediction (P1) is provided only to 
guide the eye, as now we have no idea about the required experimental 
corrections.}
\label{fig:exp-com-1}
\end{figure}

The first measurements of $^{6}$Be decay correlations were made in 
Ref.~\cite{gee77}, see Fig.~\ref{fig:exp-com-1}(a). They determined the energy
spectrum of $\alpha $-particles reconstructed in the $^{6}$Be c.m.\ frame.
For $^{6}$Be g.s.\ events, this spectrum is the same as the correlation
spectrum between two protons. The authors could not fit the data using
simplistic decay scenarios (phase volume, diproton decay, simultaneous
emission of $p$-wave protons) and concluded: ``...no
incoherent sum of the processes considered here will fit the data. Perhaps a
full three-body computation is necessary to understand the energy
spectrum.''

This ground-state decay, as well as decays of the $2^{+}$, $T=0$ states of
the $A$=6 isobar, was further investigated in the series of works by the
Kurchatov Institute group \cite[and Refs.\ therein]{boc87,boc89,boc92}, see
Fig.~\ref{fig:exp-com-1}(b). They developed a method of analyzing the $p$-$p$
correlations in the framework of a three-body partial-wave decomposition and
applied this to the three-body decays of light nuclei \cite{dan87,boc89}. In
particular, the first kinematically complete study of $^{6}$Be proved the
existence of three-particle $p$+$p$+$\alpha $ correlations with $S(p$-$p)=1$
and $S(p$-$p)=0$ \cite{boc89,boc92} which matched the three-body components
found theoretically in the $p$-shell structure of $^{6}$Be \cite{dan91}. One
of the important result for $^{6}$Be g.s.\ was the realization that $S(p$-$%
p)=0$ and $S(p$-$p)=1$ components of the WF should produce very different
correlation patterns. The presence of an ``admixture'' of $S(p$-$p)=1$
component to the WF was demonstrated by an experiment performed with special
kinematics. In these works, the concept of ``democratic decay'' was coined.
This describes the specific decay mode for three-body systems, when the
events are not highly focused in narrow kinematical regions, but are
distributed broadly (``democracy''  among different kinematical regions).
``Democratic decay'' is now a popular term for this class of phenomena, but
the correlations in $^{6}$Be decay have never been studied since that time.
The spectra shown in Figs.~\ref{fig:exp-com-1} (a) and (b) are not in
complete agreement with each other. Furthermore, there are large statistical
uncertainties and the geometry of experiments may cause cuts in kinematical
space which make comparison the theory difficult. It is clear a modern
experiment on $^{6}$Be decay was needed.


\section{Experiment}



\subsection{Experimental Method}


The Texas A\&M University K500 cyclotron facility was used to produce a
200~pnA beam of $^{10}$B at $E/A=15.0$ MeV. This primary beam impinged on a
hydrogen gas cell held at a pressure of 2 atmospheres and kept at
liquid-nitrogen temperature. A secondary beam of $E/A=10.7$~MeV $^{10}$C was
produced through the $^{10}$B$(p,n)^{10}$C reaction and separated from other
reaction products using the MARS\ spectrometer \cite{Tribble89}. This
secondary beam, with intensity of $2\times 10^{5}$~s$^{-1}$, purity of $99.5\%$, 
an energy spread of $3\%$, and a spot size of $3.5\times 3.5$ mm
was inelastically excited due to interactions with 14.1~mg/cm$^{2}$ Be and
13.4~mg/cm$^{2}$ C targets. Ground-state $^{6}$Be fragments were created
from the $\alpha $ decay of these excited $^{10}$C particles. Following the
decay of the $^{6}$Be g.s.\ fragment, the final exit channel is $2p$+$2\alpha $.

The four decay products were detected in an array of four Si $E$-$\Delta E$
telescopes located in a plane 14~cm downstream of the target. The
telescopes, part of the HIRA array \cite{Wallace07}, consisted of a 65~$\mu $m 
thick, single-sided Si-strip $\Delta E$ detector followed by a 1.5~mm
thick, double-sided Si strip $E$ detector. All Si detectors were $6.4 \times
6.4$~cm in area with their position-sensitive faces divided into 32 strips.
The telescopes were positioned in a square arrangement with each telescope
offset from its neighbor to produce a small, central, square hole through
which the unscattered beam passed. With this arrangement, the angular range
from $\theta =1.3$ to $7.7^{\circ }$ was covered. More details of the
experimental arrange can be found in Ref.~\cite{Mercurio08}.


\subsection{Monte Carlo Simulations}


\begin{figure}[tbp]
\includegraphics*[ scale=.4]{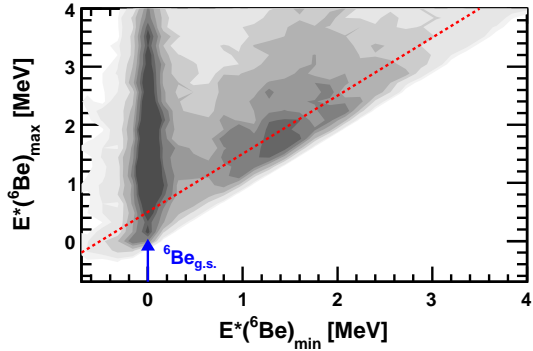}
\caption{Contour plot showing the distribution of the two possible $^{6}$Be
excitation energies that can be associated with the two $2p$+$\alpha$
subevents ordered by the maximum and minimum values. The dashed line
indicates the threshold for which correct identification of the $2p$+$\alpha$ 
subevent associated with $^{6}$Be decay is achieved in the simulations. The 
ridge associated with $^{6}$Be g.s.\ decay is indicated by the arrow.}
\label{fig:Be6_prep}
\end{figure}

Monte Carlo simulations of the experiment were performed in order to
determine the experiment bias and to understand the effects of the gates
applied to remove unwanted $2p$+$2\alpha$ events. The simulations included
the $\alpha$ decay of the parent $^{10}$C fragments and the correlations
between the $^{6}$Be decay products are sampled according to the theory of
Sect.~\ref{sec:theory}. The effects of energy loss and small-angle
scattering of all the decay products were considered following 
Refs.~\cite{Ziegler85,Anne88}.

Simulated events were passed through a detector filter and the effects of
the position and energy resolution of the detector were added. The
``detected'' simulated events were subsequently analyzed in the same manner
as the experimental data. The velocity, excitation-energy, and angular
distributions of the parent $^{10}$C states were chosen such that the
secondary distributions that passed the detector filter were consistent
with the experimental results. Similar simulations for other decay modes
were found to reproduce the experimental resolution \cite{Mercurio08}.


\subsection{Event Selection}


\begin{figure}[tbp]
\includegraphics*[ scale=.4]{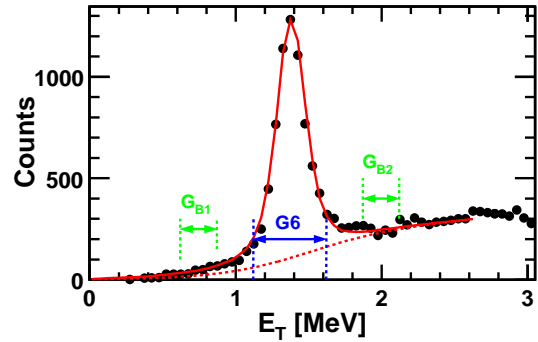}
\caption{(Color online) Experimental distribution of $E_{T}$ for selected
events is shown by the data points. The solid curve indicates the
distribution predicted by the Monte Carlo simulation with the addition of a
smooth background (dashed curve). The $G6$ gate used to select $^{6}$Be
g.s.\ events and the two gates ($G_{B1}$, $G_{B2}$), used to estimate the
background under the peak, are indicated.}
\label{fig:ET6Be}
\end{figure}

Apart from $\alpha $-$^{6}$Be g.s.\ decay, these are many other $^{10}$C
decay modes that lead to the $2p$+$2\alpha$ exit channel and thus the
detected events must be suitably gated to remove these unwanted decays. Of
particular importance is the rejection of the large yield of decays where
the $^{10}$C fragments undergoes two-proton decay (either sequential through
$^{9}$B or prompt) leading to the creation of an $^{8}$Be g.s.\  
\cite{Mercurio08}. These events can readily be identified from the correlations
between the two $\alpha $ particles. The distribution of relative energy 
($E_{rel}^{\alpha \alpha }$) between the two $\alpha $ particles contains a
strong, narrow peak corresponding to $^{8}$Be g.s. decay \cite{Mercurio08}.
This peak has a FWHM of 38~keV and sits on a negligible background 
\cite{Mercurio08} thus allowing for a clean rejection of these events with the
gate $E_{rel}^{\alpha \alpha }<0.2$~MeV.\ Our Monte Carlo simulations
suggests this gate has essentially no significant effect on true
$\alpha $-$^{6}$Be g.s.\ decays with only $0.01\%$ of detected events being 
rejected.

\begin{figure}[tbp]
\includegraphics*[ scale=.4]{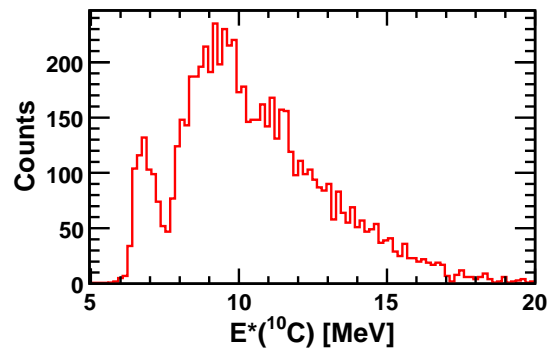}
\caption{(Color online) Experimental distribution of $^{10}$C excitation
energy for $\protect\alpha$-$^6$Be events selected in this study.}
\label{fig:ExDist}
\end{figure}

The remaining events have contributions from $\alpha $-$^{6}$Be g.s.\ and 
$p$-$^{9}$B ($E^{\ast }$=2.43 MeV) decays \cite{Mercurio08}. The latter $^{9}$B
excited state does not decay through $^{8}$Be g.s.\ but undergoes a
three-body decay like the $^{6}$Be ground state. For both of these decays modes,
there is a difficultly is trying to find the intermediate state (either $^{6}$Be 
or $^{9}$B) as there are two possible ways to construct this
fragment from the detected $2p$+$2\alpha $ exit channel. Let us concentrate
on the $^{6}$Be g.s.\ fragments first where we must determine
which of the two detected $\alpha $ particles was the one initially emitted
from the $^{10}$C parent and which was produced in the decay of $^{6}$Be. To
this end, the $^{6}$Be excitation energy for the two ways of constructing
the $^{6}$Be fragment are determined and ordered according to their maximum
and minimum values; $E^{\ast }(^{6}$Be$)_{\max }$ and $E^{\ast 
}(^{6}$Be$)_{\min}$. A two dimensional plot of these two excitation
energies is shown in Fig.~\ref{fig:Be6_prep}. A prominent ridge centered
around $E^{\ast }(^{6}$Be$)_{\min }=0$ corresponding to $^{6}$Be g.s.\ decay
is clearly visible. For those events in this ridge structure, the
identification of which $\alpha $ particles was produced in $^{6}$Be decay
is clearly the one associated with $E^{\ast }(^{6}$Be$)_{\min }$ when $E^{\ast 
}(^{6}$Be$)_{\max }\gg E^{\ast }(^{6}$Be$)_{\min }$. However when $E^{\ast 
}(^{6}$Be$)_{\max }\sim E^{\ast }(^{6}$Be$)_{\min }$ the Monte Carlo
simulations indicate that misidentifications will occur. These simulations
suggests that for $E^{\ast }(^{6}$Be$)_{\max }-E^{\ast }(^{6}$Be$)_{\min
}=0.5$~MeV, the probability of misidentifying the $\alpha $ particles is 
$0.03\%$. This condition is indicated in Fig.~\ref{fig:Be6_prep} by the
dashed line and only events above this line were used in the subsequent
analysis of the experimental data. One can see from Fig.~\ref{fig:Be6_prep}
that this condition does not significantly cut into the ridge structure and
the Monte Carlo simulations suggests we lose $4.7\%$ of the remaining $\alpha 
$-$^{6}$Be g.s.\ events with this gate.

The remaining ridge structure still sits on a background. Part of this
background can be traced to $^{10}$C$\rightarrow p$+$^{9}$B($E^{\ast }=2.43$
MeV) decays. These events can be identified from $E^{\ast }(^{9}$B$)_{\max }$
and $E^{\ast }(^{9}$B$)_{\min }$ information in a manner similar to the $\alpha 
$-$^{6}$Be g.s.\ events. A ridge structure also is evident in this
case and it also sits on an non-negligible background, which in turn has
contributions from $\alpha $-$^{6}$Be g.s.\ decay. Although one cannot
completely separate all $p$-$^{9}$B and $\alpha $-$^{6}$Be events, we do
reject events in the $E^{\ast }(^{9}$B$)_{\min }$ ridge structure. This
results in a slightly diminished yield of true $\alpha $-$^{6}$Be g.s.\
events, but more importantly, it reduces the relative background under the 
$^{6}$Be ridge structure shown in Fig.~\ref{fig:Be6_prep}. The Monte Carlo
simulations suggests only $2.7\%$ of the remaining true $\alpha $-$^{6}$Be
g.s.\ events were rejected by this condition.

\begin{figure}[tbp]
\includegraphics*[ width=.47\textwidth]{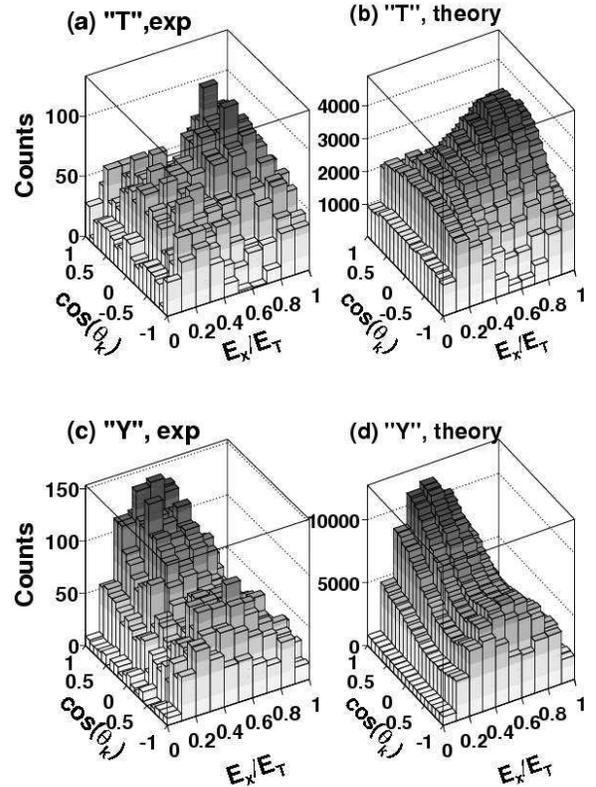}
\caption{Comparison of (a,c) experimental (exp) and (b,d) theoretical
correlations for $^{6}$Be g.s.\ decay presented in both the (a,b) ``T'' and
(c,d) ``Y'' Jacobi systems. The theoretical distributions include the
effects of the detector bias and resolution.}
\label{fig:lego}
\end{figure}

The distribution of $E_{T}$ for the final selection of events is shown in 
Fig.~\ref{fig:ET6Be} by the data points. The FWHM width of the peak associated 
with $^{6}$Be g.s.\ is 220 keV which is larger than the intrinsic value of 
$\Gamma=92$~keV due to detector resolution. The solid curve indicates the 
simulated distribution after a smooth background contribution (dashed curve) is 
added. This simulated distribution reproduces the experimental results quite 
well confirming that the Monte Carlo simulations correctly model the 
experimental resolution. Figure~\ref{fig:ET6Be} also shows the gate $G6$ used to 
select $^{6}$Be g.s.\ fragments and the two gates, $G_{B1}$ and $G_{B2}$ which, 
when combined, were used to estimate the background in the $G6$ gate. In all
subsequent results, this background has been subtracted.

The excitation-energy distribution of $^{10}$C fragments associated with the
selected events is shown in Fig.~\ref{fig:ExDist}. There is localized
strength around $E^{\ast }$($^{10}$C$)=7$ MeV and a continuous distribution
up to approximately 15~MeV. Thus many $^{10}$C excited states are
contributing to the detected $^{6}$Be g.s.\ yield.


\section{Comparison of theory and experiment}


\begin{figure}[tbp]
\includegraphics*[ scale=.43]{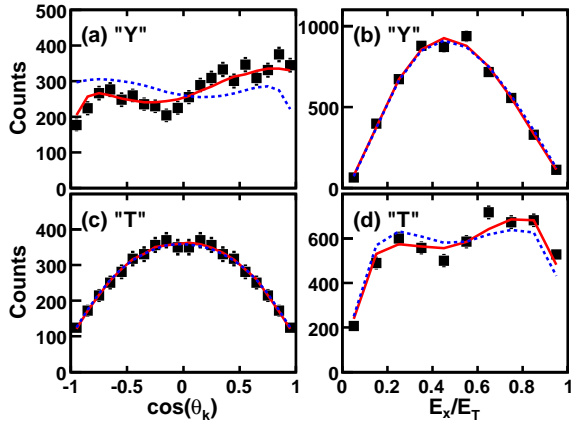}
\caption{(Color online) Comparison of the experimental (data points) and
predicted (curves) distributions of $E_{x}/E_{T}$ in the ``T'' (c), (d) and
``Y'' (a), (b) Jacobi systems. The blue dashed curves show the primary
predicted distributions while the red solid curves include the effect of the
detector bias and resolution. The theoretical results were obtained with the
P2 potential.}
\label{fig:all2}
\end{figure}

\begin{figure}[tbp]
\includegraphics*[ scale=.43]{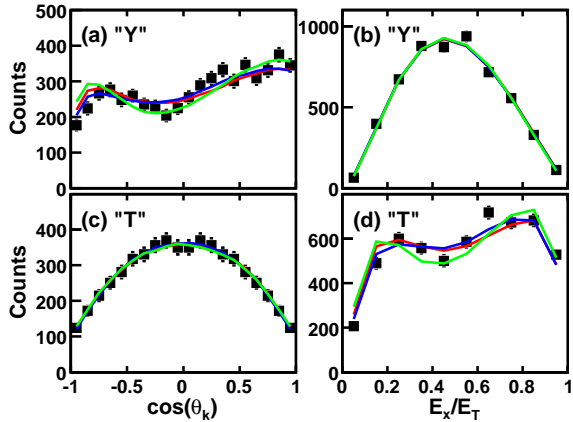}
\caption{(Color online) Comparison of the experimental (data points) and
different predicted (curves) distributions of $E_{x}/E_{T}$ in the ``T''
(c), (d) and ``Y'' (a), (b) Jacobi systems. The red, blue, and green curves
correspond to P1, P2, and P3 potential sets respectively. The effect of the
detector bias and resolution is included for the theoretical curves.}
\label{fig:all4}
\end{figure}

Comparisons of experimental and predicted correlations in both the ``T'' and
``Y'' Jacobi systems are shown in Fig.~\ref{fig:lego}. The experimental
results [Figs.~\ref{fig:lego}(a) and \ref{fig:lego}(c)] has been background
subtracted and, for the predicted distributions [Figs.~\ref{fig:lego}(b) and 
\ref{fig:lego}(d)], the effects of the detector resolution and bias has been
incorporated via the Monte Carlo simulations. In this and subsequent plots,
the simulated results has been normalized to the same number of counts as
for the experiment data. In determining the Jacobi coordinates, there are
two ways of choosing the order of the proton. For the experimental events,
Jacobi coordinates were determined for both of these ways and thus each
event contributes two counts to the spectra. For ``T'' system, this forces
the $\cos \left( \theta _{k}\right) $ distribution to be symmetrized around 
$\cos \left( \theta _{k}\right) $=0. General overall agreement between theory
and experiment is found, although statistical fluctuations are the limiting
factor for the experimental data.

\begin{figure}[t]
\includegraphics*[ width=.48\textwidth,clip]{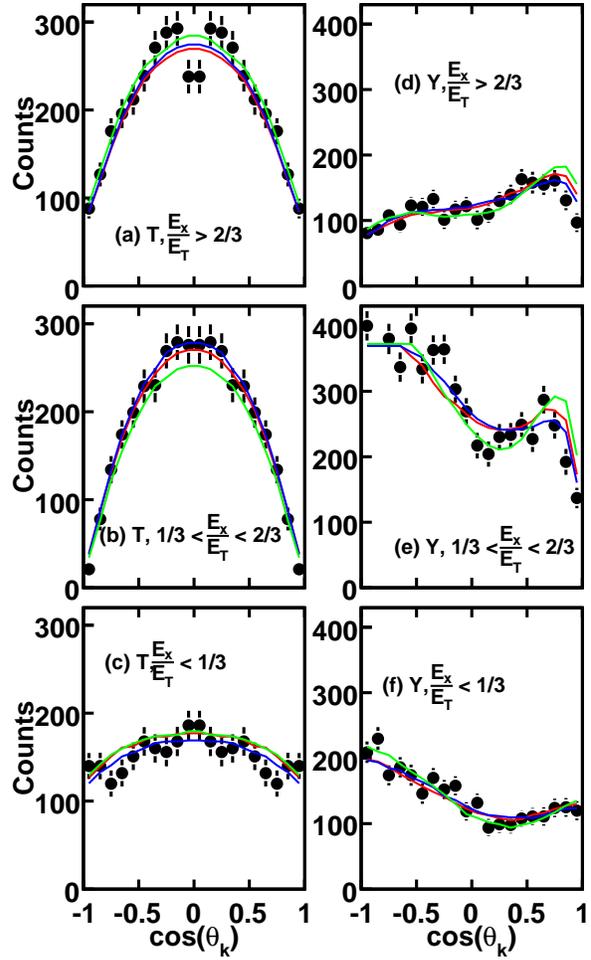}
\caption{(Color online) Comparison of experimental (data points) and
predicted (curves) $\cos (\protect\theta _{k})$ distributions in the ``T''
(left) and ``Y'' (right) Jacobi systems for the indicated gates on $%
E_x/E_{T} $ parameter. The red, blue, and green curves correspond to P1, P2,
and P3 potential sets respectively. The effect of the detector bias and
resolution is included.}
\label{fig:cos}
\end{figure}

To allow for a more detailed comparison, we compare projections of the
correlations on both the $E_{x}/E_{T}$ and $\cos \left( \theta _{k}\right) $
axes in Fig~\ref{fig:all2}. The experimental data are indicated by the data
points while the dashed and solid curves show the predictions before and
after the simulated bias of the experimental apparatus is included.
Interestingly, the ``soft'' observables (energy distribution in the ``T''
system and the angular distribution in the ``Y'' system) which have the most
sensitivity to the ingredients of the theoretical calculations and its
numerical implementation also have the largest bias induced by the detector
apparatus. The other projected distributions (angular distribution in ``T''
and energy distribution in ``Y'') are practically unaffected the detector
response.

The same comparison of theory and data for all three potentials P1-P3 is
shown in Fig.~\ref{fig:all4}. All three sets of predictions reproduce the
experimental data reasonable well. To highlight more details of the
correlations, we show the $\cos \left( \theta _{k}\right) $ distributions
gated on three equal region of $E_{x}/E_{T}$ in Fig.~\ref{fig:cos} for the
``T'' and ``Y'' Jacobi systems. Reasonable agreement between the experiment
(data points) and the three calculations (curves) is also found, although
the P1 and P2 calculation are somewhat better. \ To quantify this, we
determine the $\chi ^{2}$ per degree of freedom\ ($\chi ^{2}/\nu $) of the
theoretical fit to the two-dimension data of Fig.~\ref{fig:lego}. These
values are listed in Table~\ref{Tbl:chi}\ for both the \textquotedblleft
T\textquotedblright\ and \textquotedblleft Y\textquotedblright\ systems. For
a good fit were need $\chi ^{2}/\nu \sim $1 and clearly both P1 and P2
satisfy this criteria. Again we find the P3 calculation is somewhat worse.

\begin{table}[tbp]
\caption{$\protect\chi^2$ per degree of freedom for fits to the complete
correlations data in the ``T'' and ``Y'' system with the three assumed
potentials}
\label{Tbl:chi}
\begin{ruledtabular}
\begin{tabular}{ccc}
potential &  ``T''   & ``Y'' \\
\hline
P1 &  1.29 & 1.25 \\
P2 &  1.17 & 1.14 \\
P3 &  1.58 & 1.45 \\
\end{tabular}
\end{ruledtabular}
\end{table}


\section{Conclusions}


The first detailed studies of the correlations from the decay of $^{6}$Be
g.s.\ are performed both experimentally and theoretically. We have found
that certain correlations (namely, energy correlation between two protons
and angular correlations in ``Y'' Jacobi
system) are quite sensitive to the details of structure and interactions. We
demonstrated that relative sensitivity of correlation patterns to the
details of the interactions is higher in the decay of $^{6}$Be compared to
the corresponding sensitivity of typical observables in $^{6}$He. We argue
that further highly detailed studies of correlations in the decay of $^{6}$Be 
could provide a better access to the properties of $A$=6 isobar (and thus
to halo properties of $^{6}$He nucleus) than the direct studies of $^{6}$He
halo properties.

Experimentally $^{6}$Be fragments are produced from the $\alpha $ decay of 
$^{10}$C excited states formed by inelastically scattering a $^{10}$C beam
off of Be and C targets. The $\alpha $+$2p$ decay products as well as the
initially emitted $\alpha $ particle were detected in a Si array with good
position and energy resolution. The experimentally measured correlations
between $^{6}$Be g.s.\ decay products and the theoretical predicts were
found to be in good agreement.


\section{Acknowledgements}


This work was supported by the U.S. Department of Energy, Division of
Nuclear Physics under grants DE-FG02-87ER-40316, DE-FG02-93ER40773, and 
DE-FG02-04ER413. L.V.G. acknowledge the support from Russian Foundation for
Basic Research grants RFBR 08-02-00892, RFBR 08-02-00089-a, and Russian Ministry 
of Industry and Science grant NS-3004.2008.2.



\begin{table*}[b]
\caption{Weights $N_i$ of the dominating components of the $^{6}$He and $^{6}$Be 
g.s.\ WFs and the partial widths of the $^{6}$Be g.s.\ WF in
percent. The results are for the Jacobi ``T'' system. The normalizations of
the $^{6}$Be components are found for integration radius 
$\protect\rho_{\text{int}}=12.5$ fm.}
\label{tab:struc-t}
\begin{ruledtabular}
\begin{tabular}[c]{lcccccccccc}
 \multicolumn{2}{c}{Quantum numbers} &   \multicolumn{3}{c}{$N_i(^6$He)} &
 \multicolumn{3}{c}{$N_i(^6$Be)} & \multicolumn{3}{c}{$\Gamma_i(^6$Be)}  \\
$i$ & $K \quad L \quad S \quad l_x \quad l_y$ & P1 & P2 & P3 & P1 & P2 & P3 & P1 
& P2 &
P3 \\
\hline
1 & $0 \quad\; 0 \quad\; 0 \quad\; 0 \quad\; 0$ & 4.32  & 4.65   & 4.27  & 6.72 
 & 7.24  & 6.65  & 50.44 &
50.77 & 41.03 \\
2 & $2 \quad\; 0 \quad\; 0 \quad\; 0 \quad\; 0$ & 78.36 & 80.73  & 79.40 & 75.71 
& 77.49 & 75.28 & 33.48 & 33.74 & 41.52 \\
3 & $2 \quad\; 1 \quad\; 1 \quad\; 1 \quad\; 1$ & 14.19 & 11.28 & 12.02 & 13.09 
& 10.60 & 11.44 & 3.89  & 3.31 & 6.15 \\
4 & $4 \quad\; 0 \quad\; 0 \quad\; 0 \quad\; 0$ & 0.03  & 0.04 & 0.02  & 0.10  & 
0.14  & 0.07  & 2.03  & 2.11 & 2.25 \\
5 & $4 \quad\; 0 \quad\; 0 \quad\; 2 \quad\; 2$ & 0.48  & 0.50 & 0.58  & 0.44  & 
0.45 & 0.53  & 6.10  &
6.48 & 4.97 \\
6 & $6 \quad\; 0 \quad\; 0 \quad\; 0 \quad\; 0$ & 0.01  & 0.02 & 0.01  & 0.02  & 
0.03  & 0.01 & 1.63  & 1.26  & 1.49 \\
7 & $6 \quad\; 0 \quad\; 0 \quad\; 2 \quad\; 2$ & 1.13  & 1.18 &  1.56 & 1.56  & 
1.60 & 2.32 & 0.67  & 0.73 & 0.78 \\
8 & $6 \quad\; 1 \quad\; 1 \quad\; 3 \quad\; 3$ & 0.57  & 0.54 &  0.75 & 0.79  & 
0.75  & 1.18 & 0.08  & 0.06 & 0.09 \\
9 & $8 \quad\; 0 \quad\; 0 \quad\; 0 \quad\; 0$ & 0.28  & 0.31 &  0.37 & 0.47  & 
0.51 & 0.66 & 0.85  & 0.69 & 0.85 \\
10 & $8 \quad\; 0 \quad\; 0 \quad\; 2 \quad\; 2$ & 0.17 & 0.17 & 0.25  & 0.28  & 
0.28 & 0.46 & 0.08  & 0.11  & 0.10 \\
11 & $8 \quad\; 0 \quad\; 0 \quad\; 4 \quad\; 4$ & 0.03 & 0.03 & 0.04  & 0.05  & 
0.05 & 0.08 & 0.37  & 0.40 & 0.32 \\
\end{tabular}
\end{ruledtabular}
\end{table*}

\end{document}